\pdfoutput=1

\documentclass[11pt,twoside,a4paper,cmspaper,final,collab]{cms-tdr}
\def\svnVersion{376724}\def\cmsCernNoTag{CERN-EP/2016-191}\def\cmsCernDate{\today}\def\cmsMessage{Published in Physical Review D as \href{http://dx.doi.org/10.1103/PhysRevD.94.112005}{\doi{10.1103/PhysRevD.94.112005.}}}

\begin{document}\cmsNoteHeader{FSQ-13-010}

\hyphenation{had-ron-i-za-tion}
\hyphenation{cal-or-i-me-ter}
\hyphenation{de-vices}
\RCS$Revision: 372353 $
\RCS$HeadURL: svn+ssh://svn.cern.ch/reps/tdr2/papers/FSQ-13-010/trunk/FSQ-13-010.tex $
\RCS$Id: FSQ-13-010.tex 372353 2016-10-31 02:18:48Z alverson $
\newlength\cmsFigWidth
\ifthenelse{\boolean{cms@external}}{\setlength\cmsFigWidth{0.98\columnwidth}}{\setlength\cmsFigWidth{0.65\textwidth}}
\newlength\cmsSmallerFigWidth
\ifthenelse{\boolean{cms@external}}{\setlength\cmsSmallerFigWidth{0.7\columnwidth}}{\setlength\cmsSmallerFigWidth{0.48\textwidth}}
\ifthenelse{\boolean{cms@external}}{\providecommand{\cmsLeft}{top}}{\providecommand{\cmsLeft}{left}}
\ifthenelse{\boolean{cms@external}}{\providecommand{\cmsRight}{bottom}}{\providecommand{\cmsRight}{right}}
\providecommand{\PX}{\ensuremath{\cmsSymbolFace{X}}\xspace}
\providecommand{\Pj}{\ensuremath{\cmsSymbolFace{j}}\xspace}
\ifthenelse{\boolean{cms@external}}{\providecommand{\NA}{\ensuremath{\cdots}}}{\providecommand{\NA}{\text{---}}}
\ifthenelse{\boolean{cms@external}}{\providecommand{\cmsTable}[1]{\relax#1}}{\providecommand{\cmsTable}[1]{\resizebox{\textwidth}{!}{#1}}}

\cmsNoteHeader{FSQ-13-010}
\title{Studies of inclusive four-jet production with two \PQb-tagged jets in proton-proton collisions at \texorpdfstring{7\TeV}{7 TeV}}

\date{\today}

\abstract{
Measurements are presented of the cross section for the production of at least four jets, of which at least two originate from \PQb quarks, in proton-proton collisions. Data collected with the CMS detector at the LHC at a center-of-mass energy of 7\TeV are used, corresponding to an integrated luminosity of 3\pbinv. The cross section is measured as a function of the jet transverse momentum for $\pt>20$\GeV, and of the jet pseudorapidity for $\abs{\eta}<2.4$ (b jets), 4.7 (untagged jets). The correlations in azimuthal angle and \pt between the jets are also studied. The inclusive cross section is measured to be  $\sigma(\Pp\Pp\to 2 \PQb + 2 \Pj  + \PX) = 69 \pm3\stat\pm24\syst\unit{nb}$. The $\eta$ and \pt distributions of the four jets and the correlations between them are well reproduced by event generators that combine perturbative QCD calculations at next-to-leading-order accuracy with contributions from parton showers and multiparton interactions.}

\hypersetup{%
pdfauthor={CMS Collaboration},%
pdftitle={Studies of inclusive four-jet production with two b-tagged jets in proton-proton collisions at 7 TeV},%
pdfsubject={CMS},%
pdfkeywords={CMS, QCD, double parton scattering, forward physics}}

\maketitle
\section{Introduction}
The production of jets with large transverse momenta (\pt) in high-energy proton-proton (pp) collisions originates from parton-parton scattering, a process well described by quantum chromodynamics (QCD), the theory of the strong interaction. The cross section is evaluated as the convolution of the partonic cross sections and the parton distribution functions (PDF) in the proton. At the CERN LHC, the inclusive cross section measured for high-\pt jet production~\cite{Aad:2011fc,Chatrchyan:2012bja,Khachatryan:2016wdh} is in good agreement with the predictions of perturbative QCD (pQCD) calculations at next-to-leading order (NLO) accuracy.

Multijet final states allow studies of further features of pQCD. While at leading order (LO) a parton pair (dijet) is produced in a single parton scattering (SPS), additional jets at lower momenta can originate from two other sources. Either they arise from additional gluon radiation from SPS, or they result from double parton scattering (DPS) processes where two different pairs of partons from the two protons collide independently. The SPS processes provide tests of higher-order pQCD calculations as well as of the parton shower evolution. The contributions from DPS processes increase with center-of-mass energies as the gluon density becomes large at low values of longitudinal momentum fraction in the protons. Experimentally, SPS and DPS contributions can be separated by exploiting the different final-state topology of the two processes. Final states arising from SPS exhibit strong azimuthal and \pt correlations among all final jets, while DPS final states predominantly have a back-to-back topology only for each of the independently produced jet pairs.
Measurements of DPS signals have been performed at different collision energies and for different channels~\cite{Abe:1993rv,Abe:1997xk,Abazov:2009gc,Abazov:2014fha,Abazov:2015nnn,Aaij:2012dz,Aaij:2015wpa}. At 7\TeV, exclusive four-jet final states have been measured by CMS~\cite{Chatrchyan:2013qza}, and \PW+dijet production has been studied by ATLAS~\cite{Aad:2013bjm} and CMS~\cite{Chatrchyan:2013xxa}. Various DPS-sensitive final states have also been measured without a direct extraction of the DPS signal by CMS~\cite{Chatrchyan:2013zja,Khachatryan:2015rra} and ATLAS~\cite{Aad:2014rua,Aad:2014kba}. The present study complements the four-jet measurement~\cite{Chatrchyan:2013qza} by selecting events with jets originating from bottom quarks (denoted as ``\PQb jets''). In a four-jet sample, the SPS and DPS contributions can be disentangled by exploiting the differences expected in the angular and momentum correlations of the measured jets, as discussed in Refs.~\cite{Berger:2009cm,Blok:2015rka,Blok:2015afa}. The requirement of \PQb jets allows grouping the four jets into two pairs according to their flavor, and selecting them with lower \pt thresholds than in the untagged case, thereby facilitating the identification of DPS contributions present in the data sample.

This paper presents a measurement of DPS-sensitive observables in heavy-flavor multijet final states. The results are compared to the predictions of various MC event generators using fixed-order NLO matrix elements, and including the contributions of parton showers and multiple parton interactions (MPI). The latter processes are needed, in particular, to describe the softer hadronic production coming from the ``underlying event'' (UE). The MC used implement the DPS component as a high-\pt extension of the modelling of MPI at \pt values of the order of 3--5\GeV~\cite{Khachatryan:2015pea}. The parameters that control the simulation of softer MPI are assumed to be the same for the generation of MPI at higher-\pt scales, \ie, of DPS processes.  This assumption is used for the predictions based on either LO or NLO matrix element calculations. The MC event generators generally simulate MPI starting from the scale corresponding to the hardest parton-parton scattering provided by the matrix element calculation. In LO event generators, such as \PYTHIA and \HERWIGpp, such a scale is the \pt of the partons participating in the hard scattering, while in NLO dijet generators, e.g., \POWHEG, or multijet generators (without NLO virtual corrections), such as \MADGRAPH, the \pt of the additional outgoing partons in the matrix element calculation is also relevant for the definition of the MPI scale.
Comparing the predictions of these generators with DPS-sensitive observables in data is an important step to validate the extrapolation from soft to hard MPI, and thereby the matching of the matrix element calculations to the simulation of the UE.

The paper is organized as follows. In Section 2, a brief detector description is presented along with details of the MC simulations. In Section 3, the event selection and analysis strategy are described, while Section 4 illustrates the corrections applied to the data and the systematic uncertainties that affect the measurement. Section 5 presents the results, which are then summarized in Section 6.

\section{The CMS detector and Monte Carlo simulation}
The central feature of the CMS apparatus is a superconducting solenoid, of 6 m internal diameter and 15 m in length, which provides a magnetic field of 3.8\unit{T}. Charged-particle trajectories are measured using silicon pixel and strip trackers that cover the pseudorapidity region $\abs{\eta}< 2.5$. An electromagnetic crystal calorimeter (ECAL), and a brass/scintillator hadron calorimeter (HCAL) surround the tracking volume and cover the region $\abs{\eta}< 3.0$. A forward quartz-fiber Cherenkov hadron calorimeter extends the coverage to $\abs{\eta} \leq 5.2$. Muons are measured in the range $\abs{\eta} < 2.4$ in gas-ionization detectors embedded in the steel flux-return yoke of the magnet. The CMS experiment uses a two-level trigger system consisting of a level-1 trigger based on custom hardware using signals from the muon detectors and the calorimeters, and a high level trigger (HLT) based on a farm of computers that have access to the full data for each event. A more detailed description of the CMS detector can be found elsewhere~\cite{Chatrchyan:2008aa}.

Samples of multijet events are produced with the following MC event generators:
\begin{itemize}
\item \PYTHIA~6.426~\cite{Sjostrand:2006za}, \PYTHIA~8.185~\cite{Sjostrand:2007gs}, and \HERWIGpp 2.5.0~\cite{Bahr:2008pv}. All of them use LO 2$\to$2 matrix elements. The \PYTHIA~6 and \PYTHIA~8 event generators simulate parton showers ordered in \pt and use the Lund string model~\cite{Andersson:1998tv} for hadronization, while \HERWIGpp assumes parton showers with radiated gluons ordered in emission angle (angular ordering), and uses a cluster fragmentation model~\cite{Webber:1983if} for hadronization. The \PYTHIA and \HERWIGpp samples are generated with transverse momentum of the outgoing partons $\hat{p}_{\mathrm{T}} > 15$\GeV. The contribution of MPI is also simulated in \PYTHIA and \HERWIGpp. The \PYTHIA~6 event generator with tune Z2*~\cite{Chatrchyan:2013gfi} uses a model~\cite{skands2007} where MPI are interleaved with parton showering. Predictions obtained with \PYTHIA~6 and \PYTHIA~8 with the CUETS1 tunes~\cite{Khachatryan:2015pea} are also considered. These use the CTEQ6L1 PDF set~\cite{Pumplin:2002vw} and include an improved set of UE parameters~\cite{Khachatryan:2015pea}. The \HERWIGpp event generator with two tunes to LHC data, UE-EE-3~\cite{Gieseke:2011na} with the MRST LO** PDF set~\cite{Thorne:2009ky,Sherstnev:2007nd} and UE-EE-5-CTEQ6L1~\cite{Seymour:2013qka} with the CTEQ6L1 PDF set, is also used for comparison. The parameters of the hadronization model are determined from LEP data for both \PYTHIA~\cite{Corke:2010yf} and \HERWIGpp~\cite{Gieseke:2011na}.
\item \POWHEG 1.0~\cite{Nason:2004rx,Frixione:2007vw} matched to the \PYTHIA~8 parton showers including a simulation of MPI. The \POWHEG event generator uses NLO dijet matrix elements implemented via 2$\to$2 and 2$\to$3 diagrams. These matrix elements include only LO effects for the four-jet configuration of the present analysis. For the hard-scattering process, the HERAPDF1.5NLO~\cite{Sarkar:2014zua} PDF set is used with a minimum $\hat{p}_\mathrm{T}$ of 5\GeV. The \PQb quarks are treated as massless in the matrix element calculation. The UE provided by \PYTHIA~8 is simulated with the CUETS1 tune, which uses the HERAPDF1.5LO~\cite{Sarkar:2014zua} PDF set and reproduces with very high precision UE and jet observables at various collision energies. Since the \POWHEG predictions contain both real and virtual corrections for the dijet matrix elements, they are used as the reference baseline in the present analysis. Therefore, the full theoretical uncertainty is provided for the \POWHEG simulation, while only the central predictions are provided for the other MC simulations.
\item \MADGRAPH~5.1.5~\cite{Alwall:2011uj} interfaced with \PYTHIA~8. The \MADGRAPH predictions use a LO multijet matrix element with up to four final-state partons, calculated with the CTEQ6L1 PDF, and a simulation of the UE provided by \PYTHIA~8 tune CUETM1~\cite{Khachatryan:2015pea}, which uses the NNPDF2.3LO PDF set~\cite{Ball:2013hta,Ball:2011uy}. The \pt sum of the four partons, $H_\mathrm{T}$, is required to be  $H_\mathrm{T}>50$\GeV, and the \PQb quarks are treated as massless. The matching scale between the matrix element calculations and the parton shower simulation is taken to be 10\GeV, within the \kt-MLM scheme~\cite{Alwall:2007fs}. Underlying event data are well described by this combination of matrix elements plus parton showers with a proper UE tune~\cite{Khachatryan:2015pea}.
\end{itemize}

The detector response is simulated in detail with the \GEANTfour package~\cite{Agostinelli:2002hh}. All simulated samples are processed and reconstructed in the same manner as collision data. The multijet final state can be mimicked by various background sources, such as Drell--Yan and $\PW$ boson production associated to jets, and top-antitop events. The size of these backgrounds is estimated with \PYTHIA~8 and found to be negligible, with a cross section in the measured phase space less than 0.5\% of that for pure QCD multijet events. Therefore, these background sources are neglected in the following.

\section{Event selection}

This analysis uses data from pp collisions at $\sqrt{s} = 7$\TeV recorded with the CMS apparatus in 2010 corresponding to an integrated luminosity of 3\pbinv. The data were collected at low luminosity (${<} 0.2\times10^{33}\percms$), and consequently with low probability of multiple pp interactions in the same bunch crossing (pileup). These running conditions correspond to a fraction of the total integrated luminosity of 36\pbinv collected in 2010.
The mean number of interactions per bunch crossing is around 1.6 for this sample, which results in small pileup effects in the measured distributions. The MC samples are reweighted to the number of interactions in the data in order to match the multiplicity of reconstructed primary vertices.

For the present study, three HLT single-jet trigger sets are analyzed: one with jet \pt threshold of 15\GeV is used for leading jets with  $20 < \pt < 50$\GeV , a second with \pt threshold of 30\GeV for leading jets with  $50 < \pt < 140$\GeV, and a third with \pt threshold of 50\GeV for leading jets with \pt above 140\GeV. In the region $20 < \pt < 80$\GeV, the trigger efficiency is less than 100\%, increasing from 45\% for leading jets with $\pt\approx20\GeV$. A correction is thus applied as a function of the leading jet \pt and $\eta$. For leading jet $\pt > 80$\GeV,  the trigger is fully efficient. The choice of such regions is a compromise between statistics and reliability of the trigger efficiency correction.

The physics objects used in this analysis are particle flow (PF) jets~\cite{CMS:2009nxa}. The PF algorithm~\cite{CMS-PAS-PFT-10-001} combines information from all relevant CMS subdetectors to identify and reconstruct all particle candidates in the event, namely leptons, photons, charged and neutral hadrons. The energy of the muons is obtained from the corresponding track momentum. Charged hadrons are reconstructed from tracks in the tracker. The energy of the electrons is determined from a combination of the track momentum at the main interaction vertex, the corresponding ECAL cluster energy, and the energy sum of all bremsstrahlung photons attached to the track. Photons and neutral hadrons are reconstructed from energy clusters in the ECAL and HCAL, respectively; only clusters far away from the extrapolated position of any track are used.
Jets are reconstructed from the four-momenta of the PF candidates with the anti-\kt algorithm~\cite{Cacciari:2008gp} with a distance parameter of 0.5. A tight quality selection~\cite{CMS-PAS-JME-09-008} is applied to suppress unphysical jets, \ie, jets resulting from noise in the ECAL and/or HCAL. Each jet is required to contain at least two PF candidates, one of which has to be a charged hadron. The jet energy fraction carried by neutral hadrons, photons, muons, and electrons must be less than 90\%. With these criteria, jets are selected with an efficiency greater than 99\% and a misidentification rate (\ie the probability of selecting fake jets, like \eg, originating from leptons or calorimeter noise) smaller than 0.5\% for jet $\pt>20\GeV$. A jet \pt correction is applied to both data and simulation to  account for the nonlinear response of the calorimeters and other instrumental effects. These corrections are based on in situ measurements using dijet, $\gamma$+jet, and \Z{}+jet data samples~\cite{Chatrchyan:2011ds}.

A primary vertex (PV) is identified by a collection of tracks measured in the tracker. If more than one PV is present, the vertex with the highest sum of the squared \pt of the tracks associated to it is selected. The selected vertex is required to be reconstructed from at least five charged-particle tracks and must satisfy a set of quality requirements, including $\abs{z_\mathrm{PV}} < 24\unit{cm}$ and $\rho_{\mathrm{PV}} < 2\unit{cm}$, where $z_{\mathrm{PV}}$ and $\rho_{\mathrm{PV}}$ are the longitudinal and transverse distances of the PV from the nominal interaction point in the CMS detector.

The \PQb jets are identified by using information on the secondary decay vertex of the \PQb hadrons, the impact parameter significance, \ie, the three-dimensional impact parameter divided by its resolution, and the tracks and jet kinematics~\cite{Chatrchyan:2012jua}, through the so-called ``combined secondary vertex'' (CSV) discriminant. A loose selection~\cite{Chatrchyan:2012jua} is used in the \PQb tagging algorithm, which gives a \PQb tagging efficiency on single jets larger than 75\% for jet $\pt>20\GeV$, with a maximum of 85\% at $\pt\approx150\GeV$, as estimated by simulation studies with the \PYTHIA~6 sample. The light-flavor (\PQu, \PQd, \PQs quark or gluon) mistag probability is 20\%, 10\% and 15\% for $\pt \approx20,$ 75 and 300\GeV, respectively, for $\abs{\eta} < 2$, increasing to 35\% for jets in the region $2.0 <\abs{\eta}< 2.4$. This loose selection provides a high-statistics sample, though with relatively few genuine \PQb jets. After requiring the two \PQb tags, the \PQb jet purity, \ie the percentage of selected events where both tagged jets originate from \PQb quarks, is about 12\% for this loose selection. The highest-\pt (leading) \PQb-tagged jet is a genuine \PQb jet in 18\% of the selected events, while the fraction of events where the second-highest-\pt (subleading) \PQb-tagged jet originates from a \PQb quark is about 14\%. There is a high degree of correlation between the purities of the leading and the subleading jets. From simulation studies, about 65\% of the selected events with a true leading \PQb jet also contain a true subleading \PQb jet. The \PQb jet purity of the medium selection for the \PQb tagging algorithm~\cite{Chatrchyan:2012jua} is 58\% for the current analysis. Since the results obtained with the medium selection are consistent with those obtained with the loose selection within the systematic uncertainties, we use the latter results, which have higher statistical accuracy.

The correction for the events with four jets that pass the selection criteria but for which the two \PQb-tagged jets are not genuine \PQb jets is performed through the unfolding procedure employed to obtain stable-particle level distributions (Sec.~4). The amount of this type of background is estimated from the purity of the measured distributions. The measurement of the \PQb jet purity is based on fits of the track counting high efficiency (TCHE) distributions~\cite{Chatrchyan:2012jua} of each \PQb-tagged jet with three different shape templates obtained from MC simulation, corresponding to the TCHE values for light-quark and gluon, charm, and bottom jet flavors. The TCHE discriminant corresponds to the second-highest impact parameter significance among all selected tracks belonging to the considered jet. The \PQb jet purities measured in the data and those in the simulation differ by 2--7\%. Scale factors ($\mathrm{SF}_{\text{\PQb-purity}}$), depending on jet \pt and $\eta$, are applied to the simulation to correct for this difference. By applying $\mathrm{SF}_{\text{\PQb-purity}}$ to the simulated events, the \PQb jet purity of the data sample passing the analysis criteria is consistent with that of the MC. Compatible results are obtained if the CSV discriminant of the \PQb tagged jets is used in the fitting procedure, instead of TCHE distributions. The \PQb jet purity of the selection is estimated in the data separately for leading and subleading \PQb-tagged jets in different bins of \pt and $\eta$.

Additional scale factors ($\mathrm{SF}_{\text{\PQb-tag}}$) are applied to the simulation in order to match the \PQb tagging efficiencies measured in data~\cite{Chatrchyan:2012jua}. They depend on the jet \pt, $\eta$, and flavor, and range between 0.9 and 1.1.

A further reweighting as a function of $\hat{p}_\mathrm{T}$ is applied to the LO generators used for data correction, in order to improve their description of the measured distributions.

Events with at least one PV and at least four jets with $\pt>20\GeV$ are selected for the analysis: two of the four jets are the two \PQb-tagged jets with highest \pt within $\abs{\eta}< 2.4$, while the other two are the remaining highest-\pt jets selected within $\abs{\eta}<4.7$ without any \PQb tagging requirement. If two or more \PQb-tagged jets are present, the two with the highest \pt are taken as the ``b quark jet pair'' (referred to as ``bottom'' hereafter). The ``untagged jet pair'' (referred to as ``light'' hereafter) is taken as the remaining two leading jets. The two different $\eta$ ranges are chosen because the absence of the tracker in the forward region does not allow \PQb jets to be identified for $\abs{\eta}> 2.4$.

\begin{figure*}[htbp]
\centering
    \includegraphics[width=0.49\textwidth]{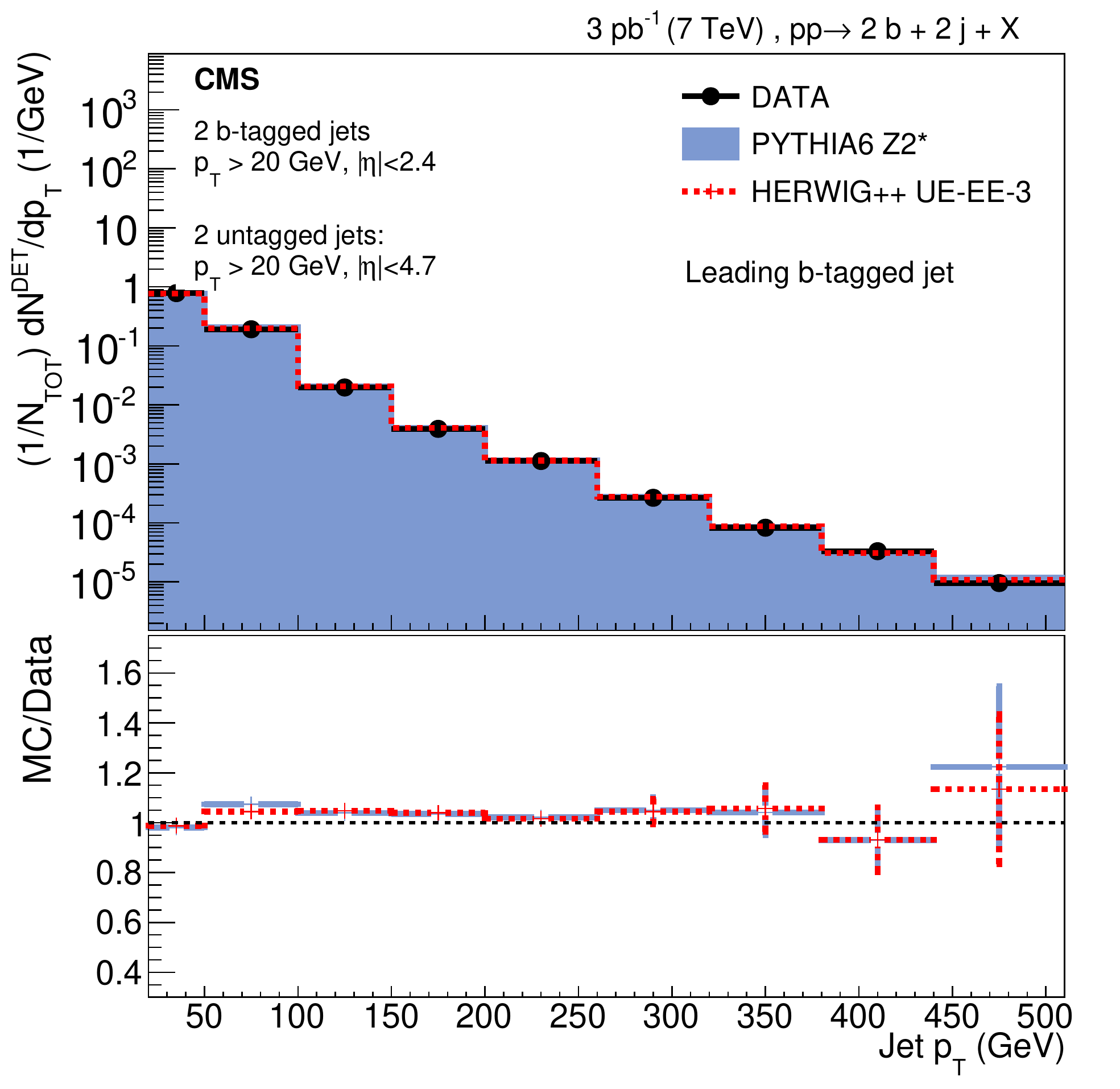}	
    \includegraphics[width=0.49\textwidth]{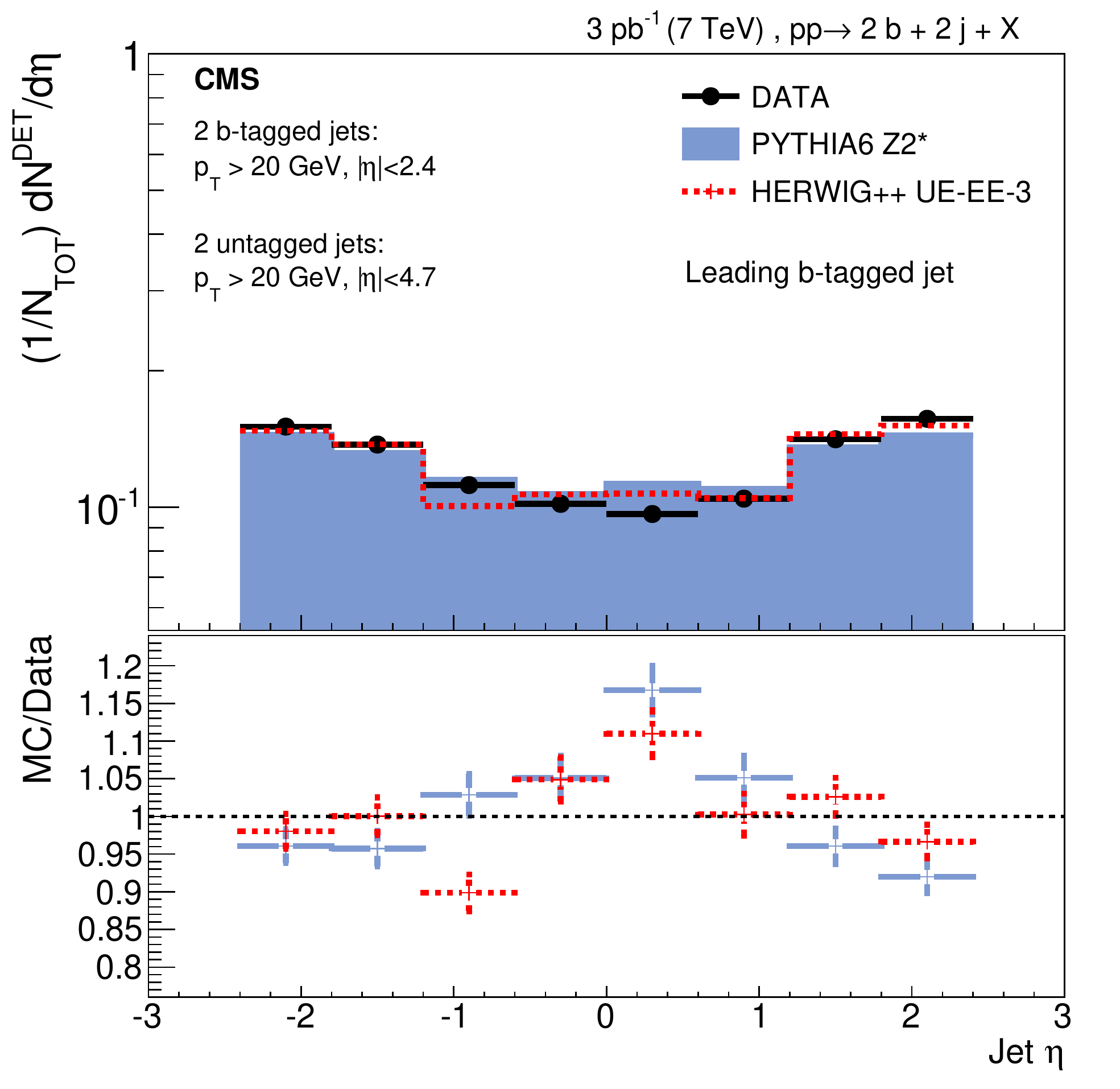}\\
    \includegraphics[width=0.49\textwidth]{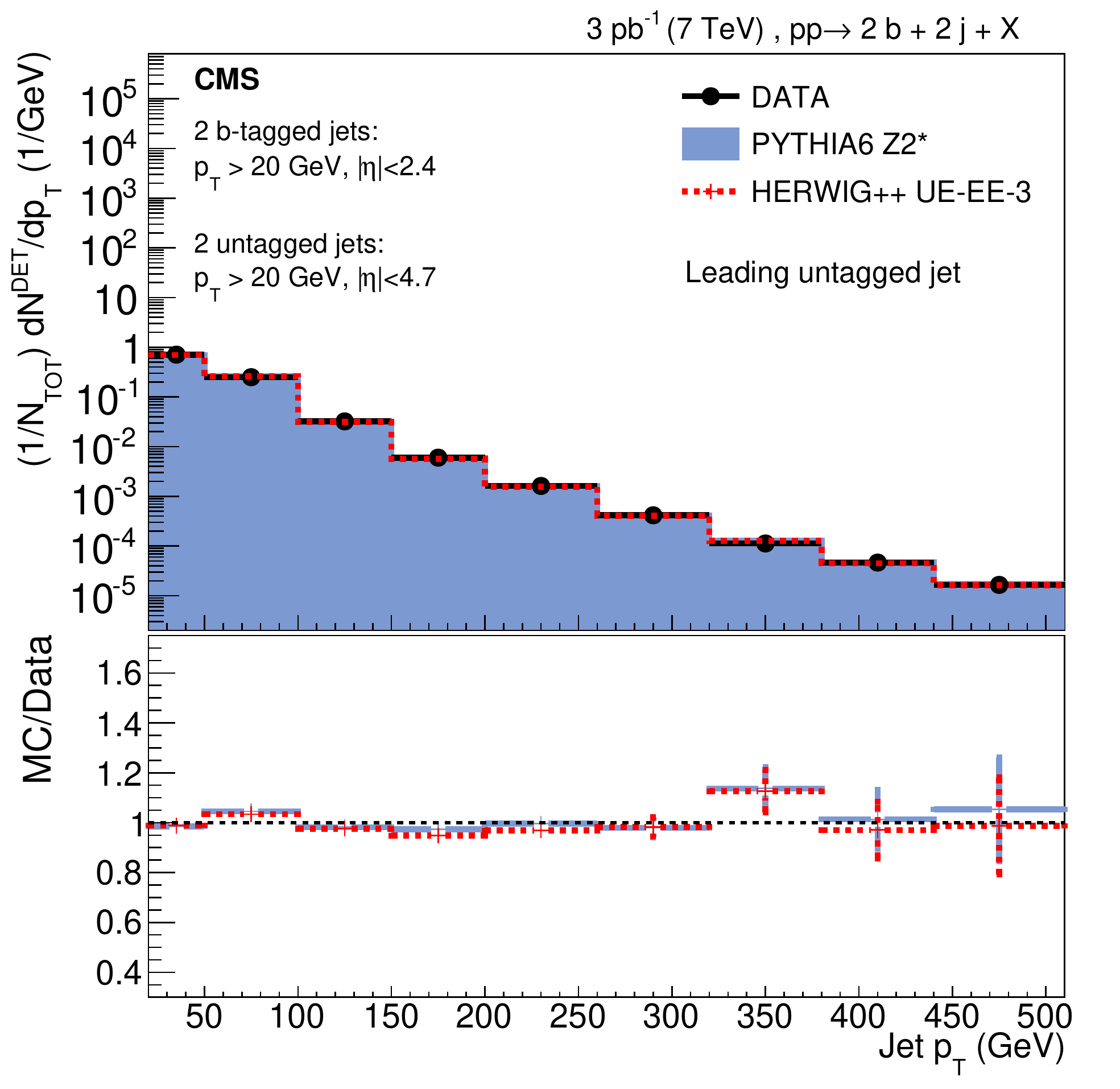}
    \includegraphics[width=0.49\textwidth]{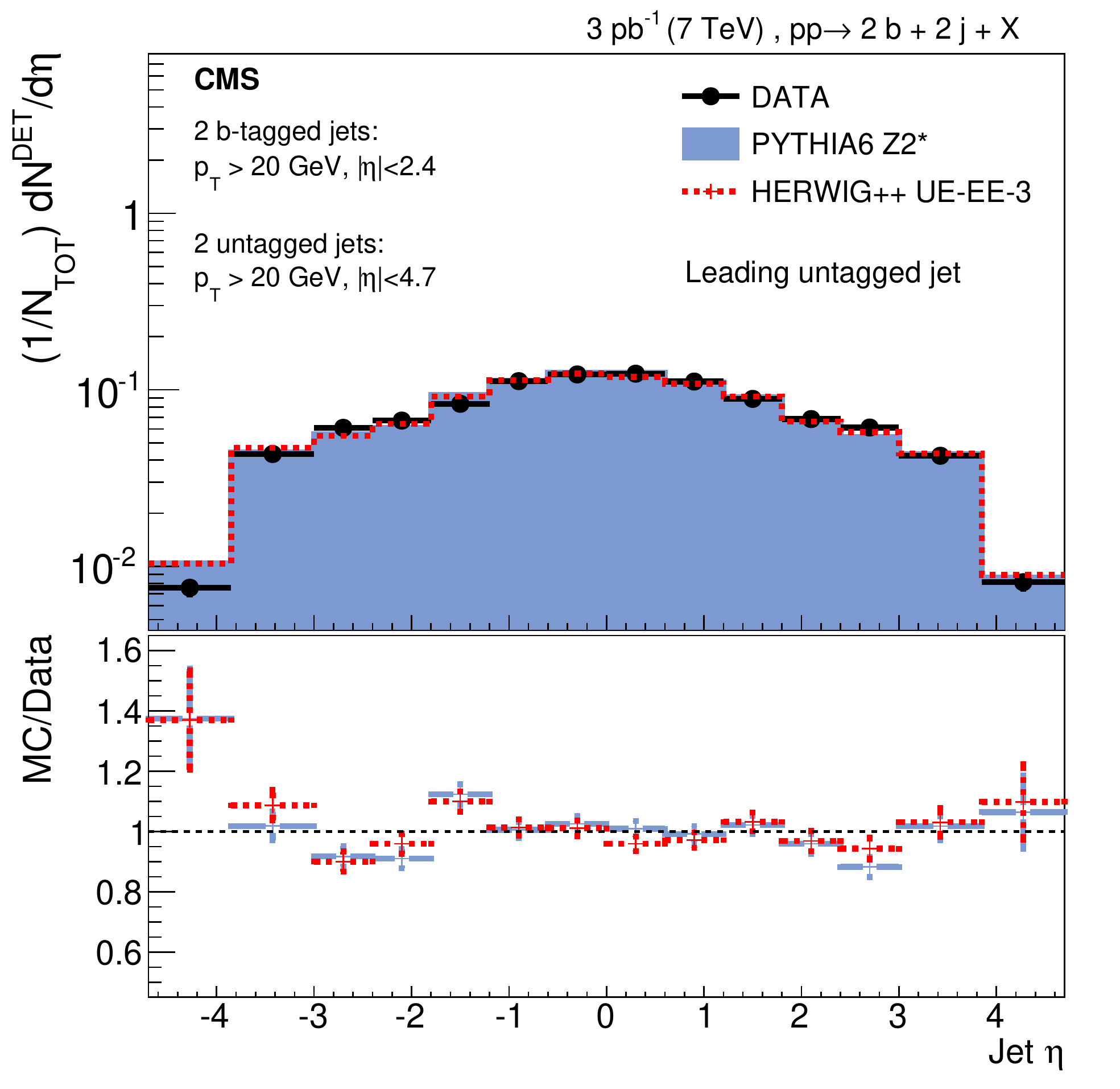}
    \caption{Uncorrected transverse momentum (left) and pseudorapidity (right) distributions of data and simulations (\PYTHIA~6 and \HERWIGpp) for the leading \PQb-tagged (top) and leading untagged (bottom) jets.  Only statistical uncertainties are shown.}
    \label{figControlDistr}
\end{figure*}

About 60\,000 events are left in the data after the offline selection described above. In Fig.~\ref{figControlDistr}, the shapes of the \pt and $\eta$ distributions of the leading \PQb-tagged and the leading untagged jet are compared to predictions of \PYTHIA~6 and \HERWIGpp, before unfolding to the stable-particle level. These shapes are well described by both MCs in the central region and over the whole range of \pt, while there are differences of up to 20--40\% for the most forward pseudorapidities ($\abs{\eta}>$ 3).

Differential cross sections (referred to as ``absolute cross sections'' hereafter) as a function of \pt and $\eta$ of each of the four jets are measured in this analysis. In addition, differential distributions normalized to the total number of selected events (referred to as ``normalized cross sections'') are measured as a function of jet correlation variables very similar to those used in the four-jet analysis of Ref.~\cite{Chatrchyan:2013qza}:
\begin{itemize}
\item the difference in azimuthal angle (in the plane transverse to the beam axis, in radians) between the jets belonging to the light-jet pair:
\begin{equation}
\Delta\phi_\text{light}=\abs{\phi_{\text{light}_1}-\phi_{\text{light}_2}};\\
\end{equation}
\item the balance in \pt of the two light jets:
\begin{equation}
\Delta^\text{rel}_\text{light}\pt = \frac{\abs{\ptvec^{\text{light}_1}+\ptvec^{\text{light}_2}}}{\abs{\ptvec^{\text{light}_1}}+\abs{\ptvec^{\text{light}_2}}};
\end{equation}
\item the azimuthal angle $\Delta$S between the two dijet pairs, defined as:
\begin{equation}
\Delta S=\arccos \left[ \frac{\ptvec(\text{bottom}_1,\text{bottom}_2)\cdot \ptvec(\text{light}_1,\text{light}_2)}{\abs{\ptvec(\text{bottom}_1,\text{bottom}_2)}\cdot \abs{\ptvec(\text{light}_1,\text{light}_2)}} \right],
\end{equation}
\end{itemize}
where $\text{bottom}_1$ ($\text{bottom}_2$) and $\text{light}_1$ ($\text{light}_2$) are the leading (subleading) jets of the bottom and light jet pairs, respectively, and $\ptvec(\text{bottom}_1,\text{bottom}_2)$ and $\ptvec(\text{light}_1,\text{light}_2)$ the momentum vectors of each pair, obtained as the vectorial sum of the momenta of the bottom and light jets, respectively.

Results of the jet correlation observables are presented as distributions normalized to the number of events measured in the selected kinematic region. Such normalized distributions are affected by smaller systematic uncertainties than the absolute cross section measurements.

\section{Corrections and systematic uncertainties}
\begin{figure*}[htbp]
\centering
    \includegraphics[width=0.49\textwidth]{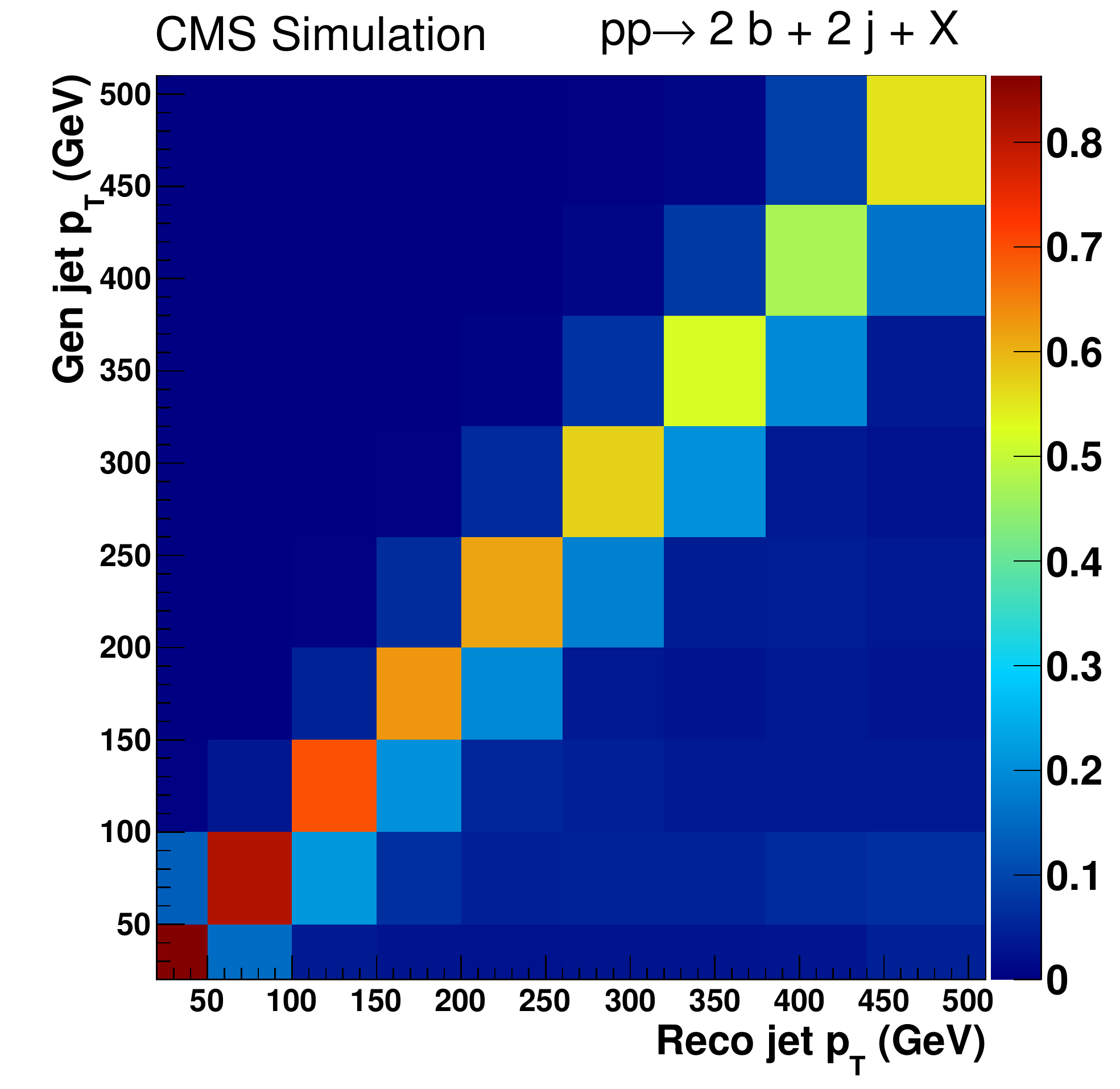}	
    \includegraphics[width=0.49\textwidth]{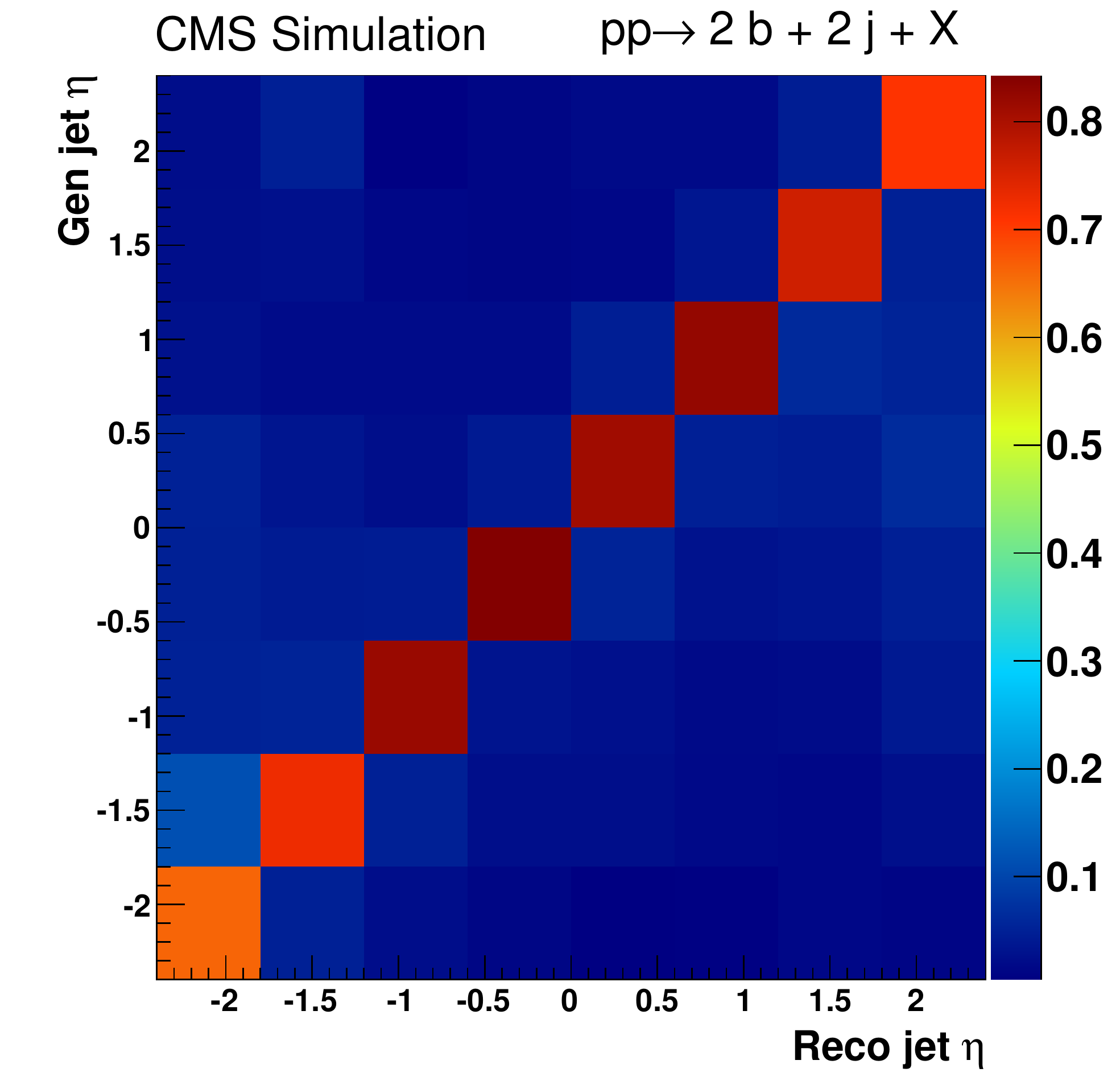}\\
    \includegraphics[width=0.49\textwidth]{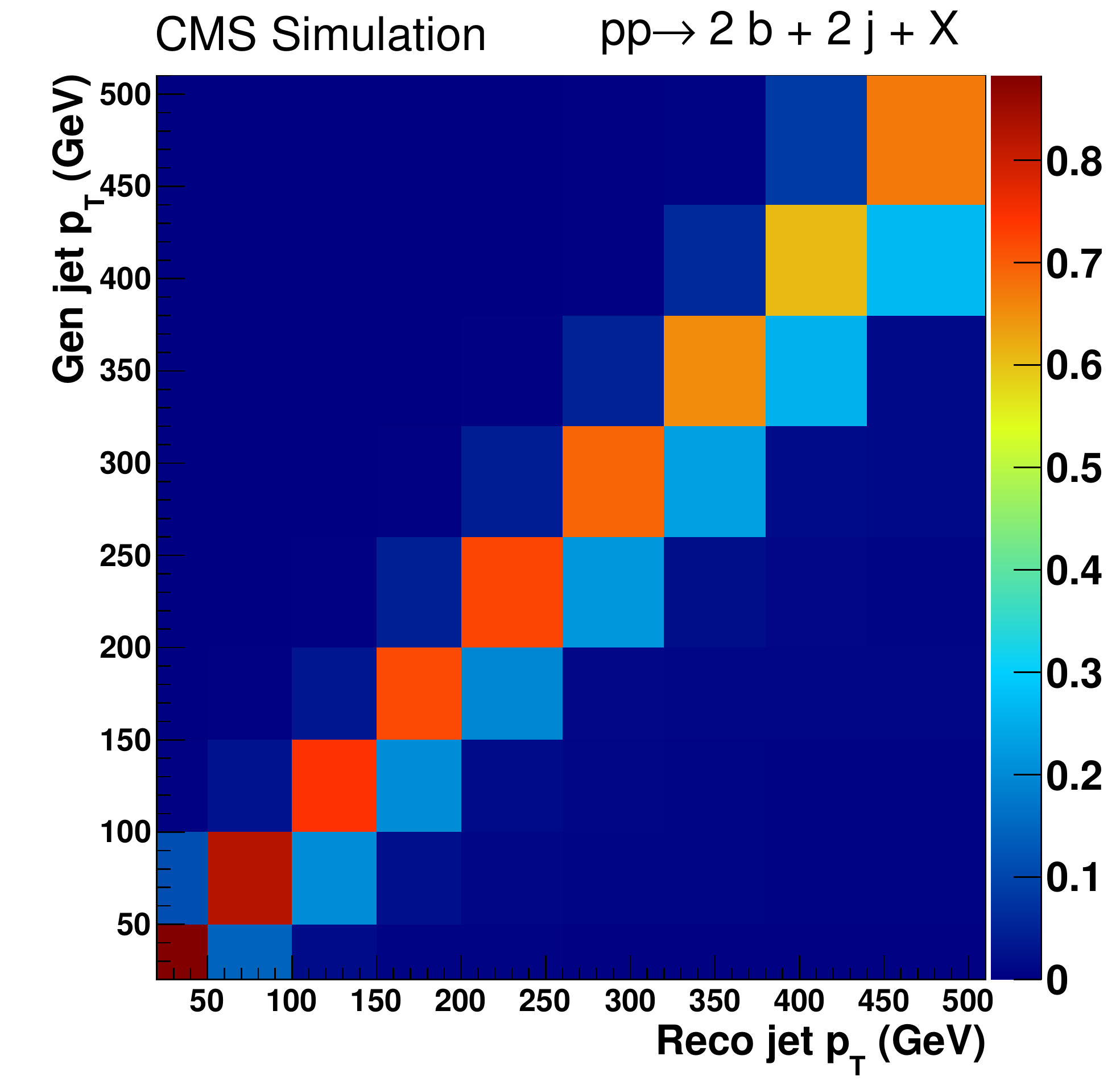}
    \includegraphics[width=0.49\textwidth]{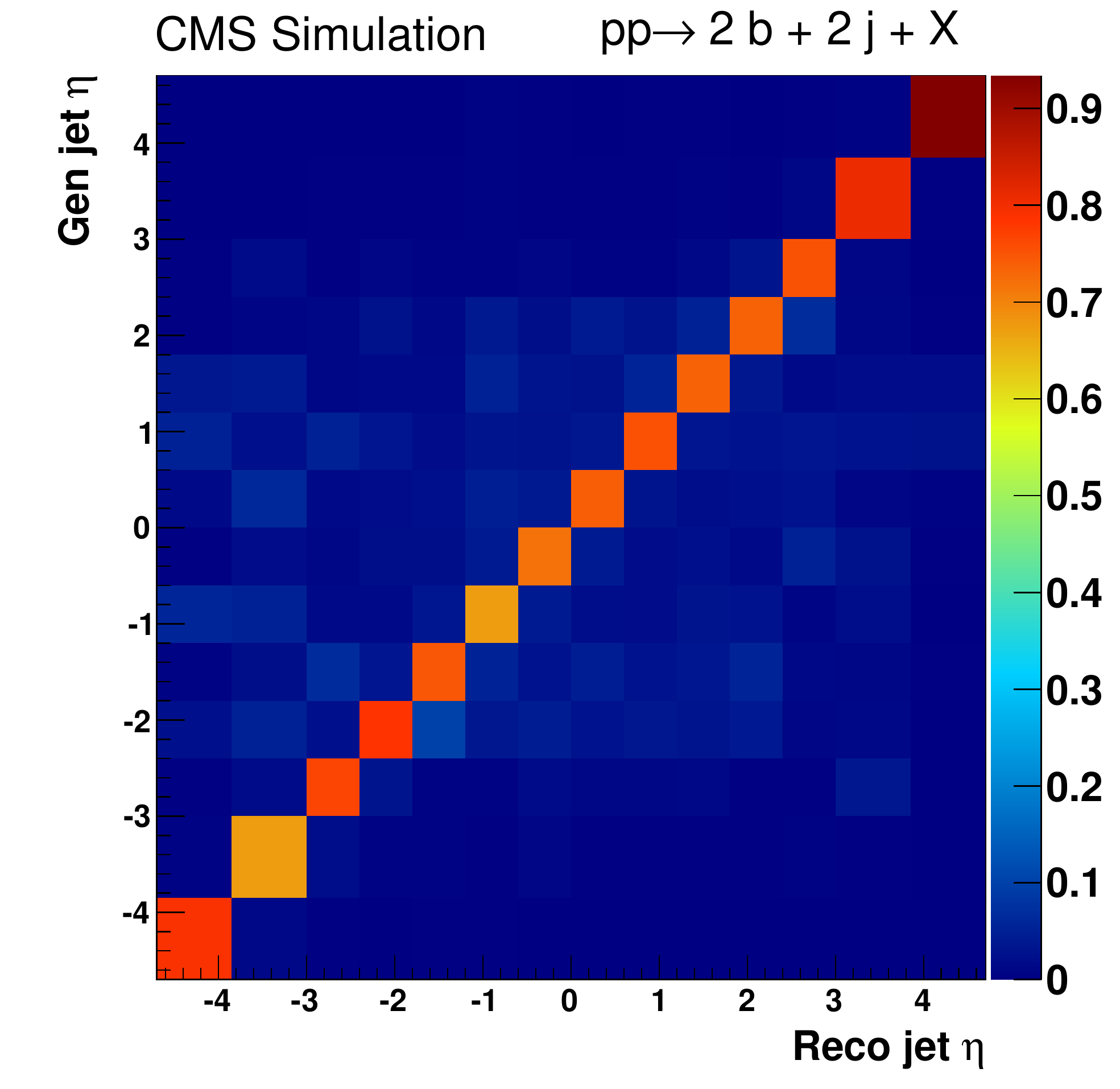}
    \caption{Response matrices obtained with the \PYTHIA~6 tune Z2* simulation for the transverse momentum (left) and pseudorapidity (right) of the leading \PQb-tagged (top) and leading untagged (bottom) jets.}
    \label{figResponseMatrix}
\end{figure*}

Particle-level distributions are inferred from the reconstructed data by correcting for selection efficiencies and detector effects. The results are corrected to particle level by applying an iterative unfolding~\cite{D'Agostini:1994zf} as implemented in the \textsc{RooUnfold} package~\cite{Adye:2011gm}. Particles are considered stable if their mean path length $c\tau$ is greater than 10\unit{mm}. MC jets are identified as ``b jets'' at the particle level if a \PQb quark is found within a cone of radius R = $\sqrt{\smash[b]{(\Delta\eta)^2+(\Delta\phi)^2}} = 0.3$ around the jet axis. The background consisting of events with four jets that pass the selection criteria but for which the \PQb-tagged jets are not genuine \PQb jets is corrected for with \PYTHIA~6 tune Z2*, after applying the $\mathrm{SF}_{\text{\PQb-tag}}$ and $\mathrm{SF}_{\text{\PQb-purity}}$ scale factors. The correlation between events selected at the reconstructed and particle levels is then studied by constructing the response matrix. The response matrix quantifies the migration probability between the particle level and reconstructed quantities, as well as the overall reconstruction efficiency. It is obtained for each observable with the \PYTHIA~6 tune Z2* sample. Diagonal terms in the response matrix correspond to particle-level quantities that are reconstructed in the same bin after detector simulation. Off-diagonal terms represent the probability of migration between bins at the particle level and bins at the reconstructed level. As an example, Fig.~\ref{figResponseMatrix} shows the response matrices for the \pt and the $\eta$ of the leading \PQb-tagged and the leading untagged jet. They exhibit a diagonal structure, with off-diagonal terms less than 30--40\%. The bin widths are larger than the detector resolution at each bin.

The response matrix obtained with \PYTHIA~6 is used for the data unfolding. As a cross check, a sample of events generated with \HERWIGpp tune UE-EE-3 is unfolded with the \PYTHIA~6 response matrix. All distributions agree with the generated ones within 9--20\%. The iterative unfolding procedure is regularized by limiting the number of iterations to a certain value for each measured distribution. The optimal number of iterations is determined by minimizing the difference between the distributions measured in the data and the ones obtained by applying backwards the detector effects to the unfolded distributions. The number of iterations ranges between 2 and 4 depending on the observable. As expected, the statistical uncertainties of the unfolded distributions are larger than those of the reconstructed data. The unfolding to particle level includes corrections for jet resolution, flavor misidentification, and pileup effects. The results are presented in the kinematic region defined in Table \ref{tablestable}.

\begin{table}[htbp]
\topcaption{Phase space for the cross section measurement.}
\centering
\begin{scotch}{l l}
At least four jets & $\pt>20$\GeV\\
Two leading \PQb jets & $\abs{\eta} < 2.4$\\
Two leading other jets & $\abs{\eta} < 4.7$\\
\end{scotch}
\label{tablestable}
\end{table}

All significant sources of systematic uncertainties are investigated and the corresponding uncertainty is calculated for each distribution. The total uncertainty is obtained by summing up the individual contributions in quadrature. The following systematic effects are considered:
\begin{description}
\item[Model dependence] the response matrix obtained with \PYTHIA~6 is used for the final correction, and the difference between this and that obtained with \HERWIGpp is taken as a measure of the model dependence of the unfolding, resulting in an uncertainty ranging from 9\% to 20\%.
\item[Jet energy scale (JES)] the momentum of the jets is varied according to the uncertainty associated with the reconstructed \pt~\cite{Chatrchyan:2011ds}. The resulting uncertainty is of the order of 20--25\% (5\%) for the absolute (normalized) cross sections.
\item[Jet energy resolution (JER)] the JER differs for data and simulation by 6--19\%~\cite{Chatrchyan:2011ds} depending on the $\eta$ range, and introduces a systematic uncertainty of 4--8\% in all results.
\item[Pileup reweighting] the effect of the pileup reweighting procedure is evaluated and found to be negligible ($<$ 0.1\%).
\item[$\PB$ tagging scale factor ($\mathrm{SF}_{\text{\PQb-tag}}$)] the values of the scale factors are varied by 10\% for each jet flavor~\cite{Chatrchyan:2012jua}. This variation results in an uncertainty of 15--18\% for absolute cross sections and of 1--2\% for the normalized ones.
\item[$\PB$ jet purity] the \PQb jet purity of the sample is evaluated by fitting separately the TCHE distribution of the leading and of the subleading \PQb-tagged jet in bins of \pt, $\eta$ and $\Delta$S. The difference between the unfolded results when using the $\mathrm{SF}_{\text{\PQb-purity}}$ obtained from the two fits is used as a systematic uncertainty, resulting in values of 10--12\% for the absolute cross sections and 1--2\% for the normalized distributions.
\item[Trigger efficiency] the trigger efficiency correction is varied within its uncertainty and the resulting corrections are applied to the data. These variations result in an uncertainty ranging from 1 to 6\%.
\item[Integrated luminosity] the systematic uncertainty on the luminosity of the 2010 data, affecting the absolute cross sections, is 4\%~\cite{CMS:2011egv}.
\end{description}

The dominant source of uncertainty is the JES, which is considered as correlated among the measured bins. The following aspects of the theoretical uncertainty affecting the \POWHEG predictions are also evaluated:

\begin{description}
\item[PDF uncertainty] the choice of the PDF set influences the theoretical predictions. The uncertainty related to the PDF is determined by generating predictions with various PDF eigenvectors. As central PDF set, the HERAPDF1.5NLO together with the \PYTHIA~6 tune CUETS1 is used.
\item[Scale uncertainty] the default renormalization and the factorization scales ($\mu_\mathrm{R}$ and $\mu_\mathrm{F}$) in the matrix element calculations are chosen to be equal to the leading jet \pt value. The uncertainty related to the $\mu_\mathrm{R}$ and $\mu_\mathrm{F}$ choices is estimated by using \POWHEG interfaced to the UE simulation provided by \PYTHIA~8 tune CUETS1-HERAPDF. Six combinations of the ($\mu_\mathrm{R}/\pt$, $\mu_\mathrm{F}/\pt$) scales: (0.5,0.5), (0.5,1), (1,0.5), (1,2), (2,1), and (2,2), are used. The scale uncertainties are evaluated by taking the envelope of the predictions obtained with the listed scale choices.
\end{description}

A summary of all the systematic effects is given in Table \ref{tableuncert}.

\begin{table*}[htbp]
\topcaption{Systematic and statistical uncertainties affecting the absolute and the normalized cross sections for each measured observable: each source of uncertainty is specified and the value is the average over all the bins of the observable. The 4\% uncertainty from the integrated luminosity is included in the total uncertainty affecting the absolute cross sections. The total uncertainty is obtained by summing the individual experimental uncertainties quadratically. The theoretical uncertainties, listed in the last two columns, affect all the predictions. The systematic uncertainties in the normalized cross sections are smaller than those for the absolute cross sections, since, among others, they are not affected by the migration effects from outside the selected phase space.}
\centering
\cmsTable{
\begin{scotch}{c c c c c c c c c | c c}
 {Measured} & {Model} & {JES} & {JER} & {$\mathrm{SF}_{\text{\PQb-tag}}$} & {$\mathrm{SF}_{\text{\PQb-purity}}$} & {Trigger} & {Stat} & {Total} & {PDF} & {Scale}\\
 {observable} &  &  & & & & {efficiency} & & {incl. int. lumi,} & & \\
\hline\\[0.2ex]
\multicolumn{11}{c}{Absolute cross sections}\\
\hline
{\PQb-tagged jet \pt} & 20\% & 25\% & 4\% & 15\% & 12\% & 6\% & 4\% & 38\% & 10\% & 10\%\\
{Untagged jet \pt} & 10\% & 25\% & 4\% & 15\% & 12\% & 6\% & 4\% & 34\% & 10\% & 10\%\\
{Jet $\abs{\eta}\leq3$} & 10\% & 25\% & 4\% & 15\% & 12\% & 5\% & 4\% & 34\% & 15\% & 10\%\\
{Jet $\abs{\eta}>3$} & 20\% & 35\% & 4\% & 15\% & 12\% & 5\% & 4\% & 45\% & 50\% & 15\% \\
\hline\\[0.2ex]
\multicolumn{11}{c}{Normalized cross sections}\\
\hline
$\Delta\phi^{\text{light}}$ & 13\% & 5\% & 1\% & 2\% & 1\% & 1\% & 4\% & 15\% & 5\% & 2\% \\
$\Delta^{\text{rel}}_{\text{light}}p_\text{T}$ & 13\% & 5\% & 7\% & 2\% & 1\% & 1\% & 4\% & 16\% & 5\% & 2\%\\
$\Delta$S & 20\% & 5\% & 10\% & 2\% & 2\% & 1\% & 4\% & 23\% & 10\% & 2\%\\
\end{scotch}}
\label{tableuncert}
\end{table*}

\section{Results}
The absolute differential cross sections are measured as a function of the jet \pt and $\eta$, along with the normalized cross sections as a function of the jet correlation variables. In Table~\ref{CrossSections}, the cross section is given, and compared to predictions from different event generators at the particle level. The \POWHEG event generator interfaced with \PYTHIA~8 tune CUETS1, referred to in the following as ``\POWHEG'', reproduces the measured cross section best. However, if the MPI simulation is switched off, the same \POWHEG predictions, referred to in the following as ``\POWHEG MPI-off'', underestimate the value of the measured cross section. All predictions are consistent with the data within uncertainties, although \MADGRAPH{}+\PYTHIA~8 tune CUETM1 (``\MADGRAPH'' in the following) tends to underestimate the data, and \PYTHIA~8 to overestimate them.

\begin{table*}[htbp]
\topcaption{Inclusive cross section for $\Pp\Pp \to 2 \PQb + 2 \Pj + \PX$ for jet $\pt>20\GeV$, with \PQb jets within $\abs{\eta}<2.4$, and the other jets within $\abs{\eta}<4.7$. The measurements are compared to the MC predictions.}
\centering
\newcolumntype{x}{D{,}{\,\pm\,}{-1}}
\begin{scotch}{l l x}
\multicolumn{1}{c}{Sample} & \multicolumn{1}{c}{PDF} & \multicolumn{1}{c}{Cross section (nb)}\\
\hline
Data & \NA & \multicolumn{1}{c}{$69\pm3\stat\pm24\syst$}\\
\hline
\POWHEG{}+\PYTHIA~8 tune CUETS1 & HERAPDF1.5 & 65,12\\
\POWHEG{}+\PYTHIA~8 tune CUETS1 MPI off & HERAPDF1.5 & 31,6\\
\hline
\PYTHIA~6 tune Z2* & CTEQ6L1 & 77,15\\
\PYTHIA~6 tune CUETS1 & CTEQ6L1 & 77,15\\
\HERWIGpp tune UE-EE-3 & MRST LO** & 44,8\\
\HERWIGpp tune UE-EE-5C & CTEQ6L1 & 47,9\\
\PYTHIA~8 tune CUETS1 & CTEQ6L1 & 96,18\\
\MADGRAPH+\PYTHIA~8 tune CUETM1 & CTEQ6L1 & 39,7\\
\end{scotch}
\label{CrossSections}
\end{table*}

\begin{figure}[htbp]
\centering
    \includegraphics[width=\cmsSmallerFigWidth]{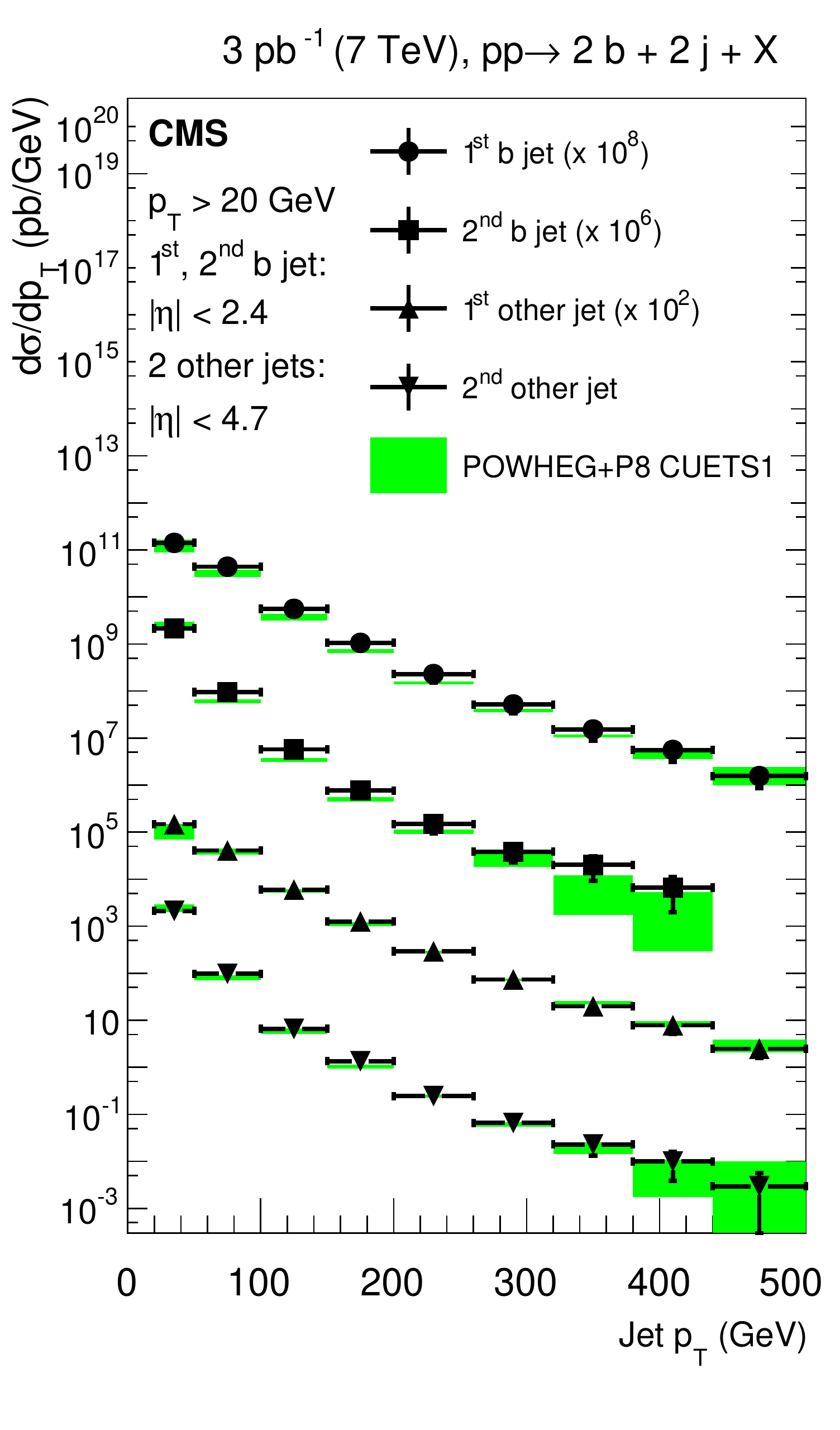}	
    \includegraphics[width=\cmsSmallerFigWidth]{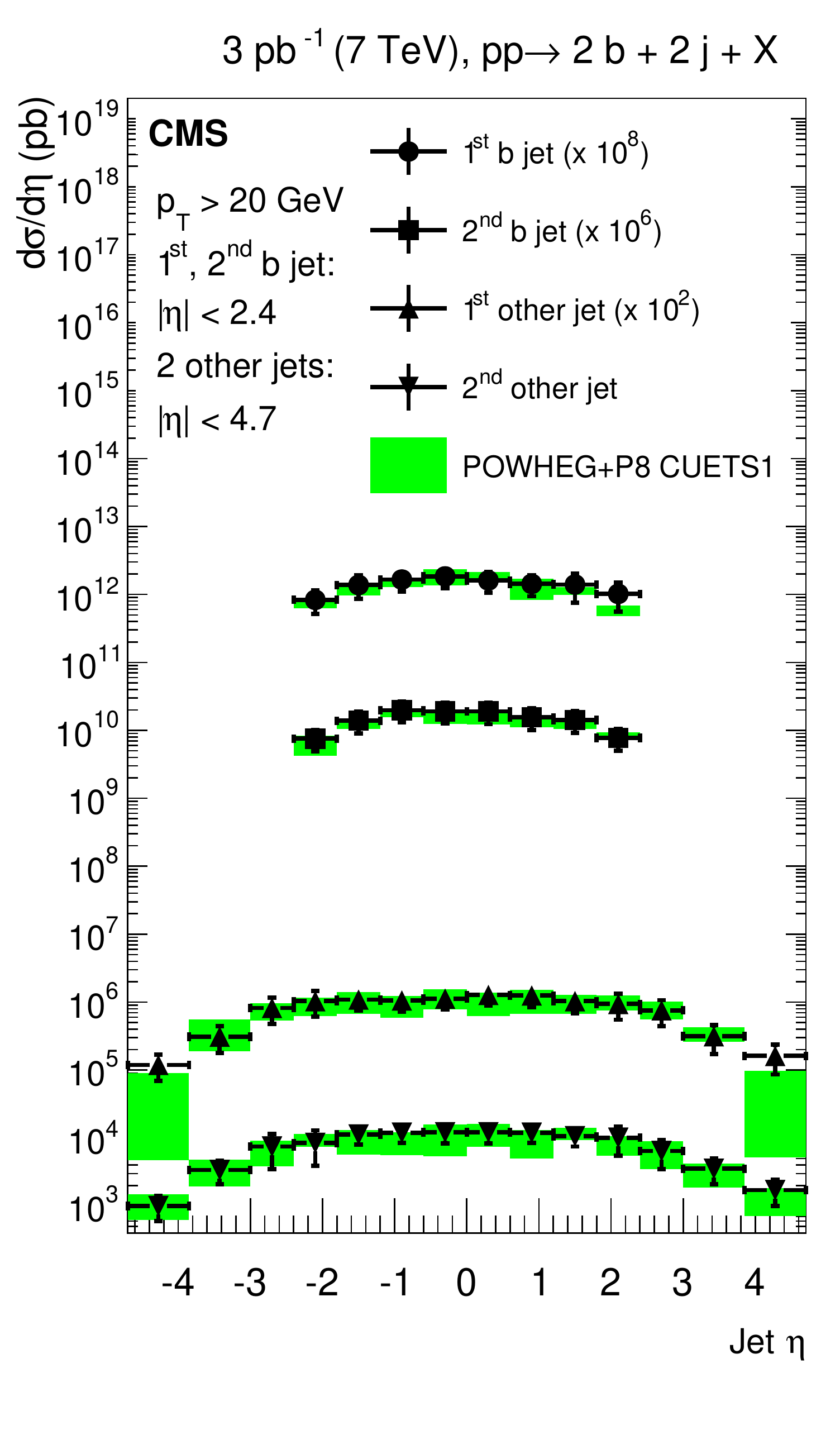}
    \caption{Differential cross sections unfolded to the particle level as a function of the jet transverse momenta \pt (left) and pseudorapidity $\eta$ (right) compared to predictions of \POWHEG{}+\PYTHIA~8 tune CUETS1. Scale factors of $10^8$, $10^6$, and $10^2$ are applied (for clarity) to the measurement of the leading, subleading, and third jet, respectively. The error bars on the data represent the total uncertainties, \ie, statistical and systematic added quadratically. The band represents the theoretical uncertainty due to the choice of the scales and PDFs.}
    \label{figPt}
\end{figure}

\begin{figure}[htbp]
\centering
   \includegraphics[width=\cmsFigWidth]{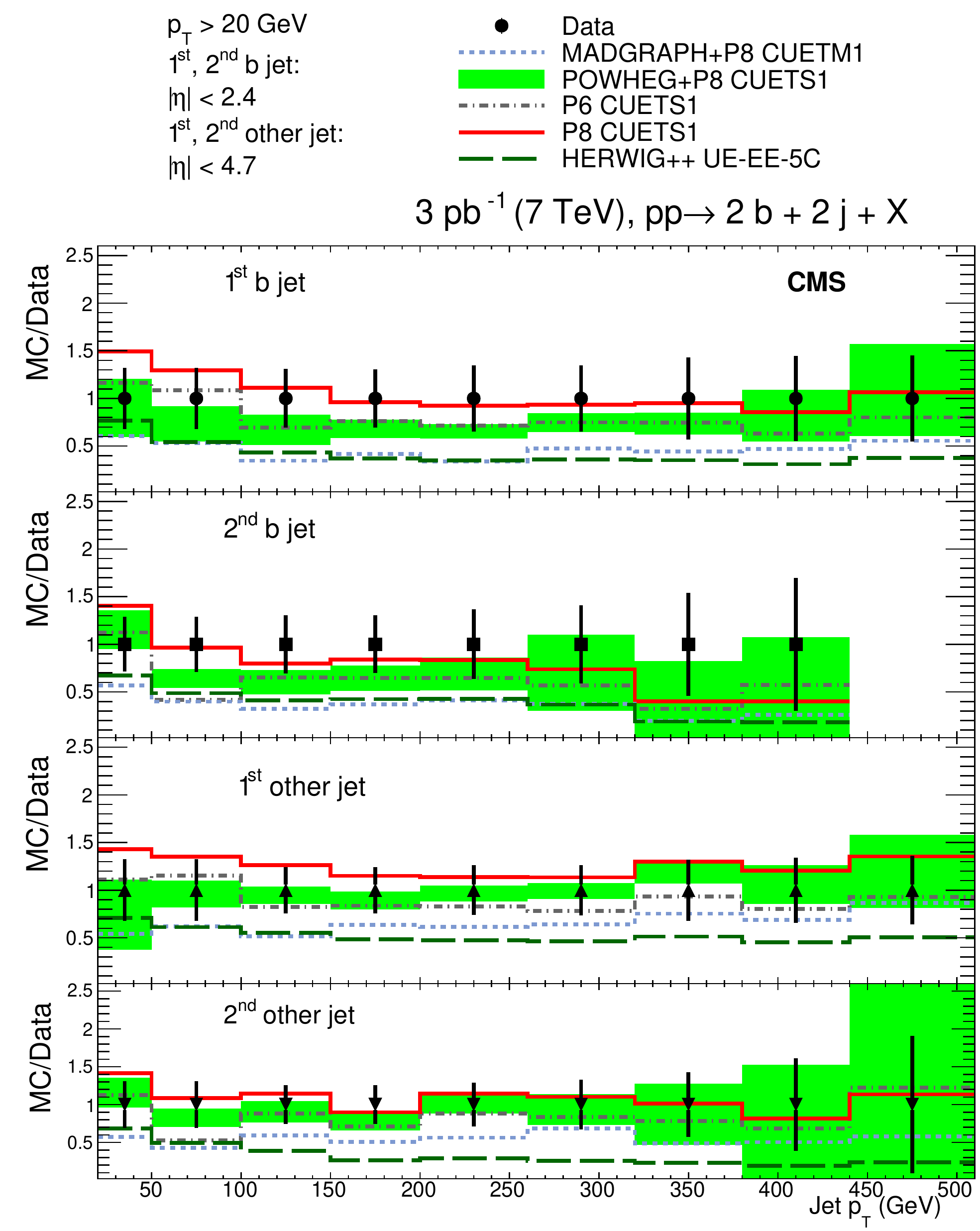}
    \caption{Ratios of the absolute cross section predictions of \POWHEG, \MADGRAPH, \PYTHIA~6 (P6), \PYTHIA~8 (P8), and \HERWIGpp over data (unfolded to the particle level) as a function of the jet transverse momenta \pt for each jet. The error bars on the data represent the total uncertainties, \ie, statistical and systematic added quadratically. Data are shown with markers at unity. The band represents the theoretical uncertainty due to the choice of the scales and PDFs (shown only around the \POWHEG ratio for clarity, but affecting all predictions in the same way).}
    \label{figEta1}
\end{figure}

\begin{figure}[htbp]
\centering
    \includegraphics[width=\cmsFigWidth]{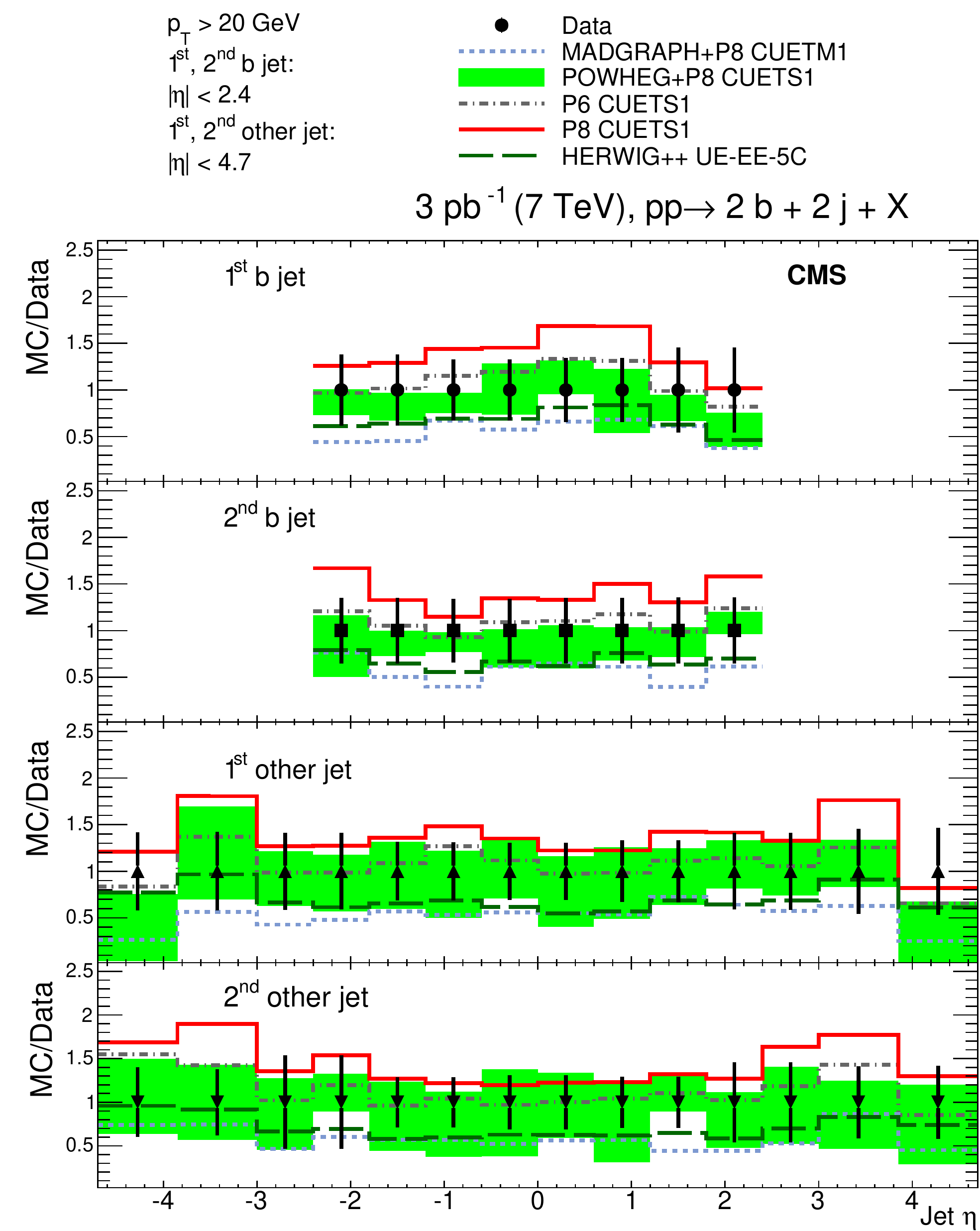}
    \caption{Ratios of the absolute cross section predictions of \POWHEG, \MADGRAPH, \PYTHIA~6 (P6), \PYTHIA~8 (P8), and \HERWIGpp over data (unfolded to the particle level) as a function of the jet pseudorapidity $\eta$ for each jet. The error bars on the data represent the total uncertainties, \ie, statistical and systematic added quadratically. Data are shown with markers at unity. The band represents the theoretical uncertainty due to the choice of the scales and PDFs (shown only around the \POWHEG ratio for clarity, but affecting all predictions in the same way).}
    \label{figEta2}
\end{figure}

In Fig.~\ref{figPt}, the absolute differential cross sections as a function of the \pt and $\eta$ of the selected jets are shown compared to predictions from \POWHEG. Figures~\ref{figEta1} and \ref{figEta2} present the same differential cross sections as ratios of theoretical predictions from various MC event generators to the data. The \POWHEG predictions reproduce very well the measurements as a function of \pt and $\eta$ of each jet, in both the central and forward regions. The other MC simulations also describe the data satisfactorily, although \HERWIGpp tune UE-EE-5C and \MADGRAPH are systematically lower than the data.
Similar conclusions about \HERWIGpp and \MADGRAPH have been already drawn for inclusive~\cite{Khachatryan:2015pea} and exclusive four-jet~\cite{Chatrchyan:2013qza} final states.

\begin{figure}[htbp]
\centering
    \includegraphics[width=\cmsSmallerFigWidth]{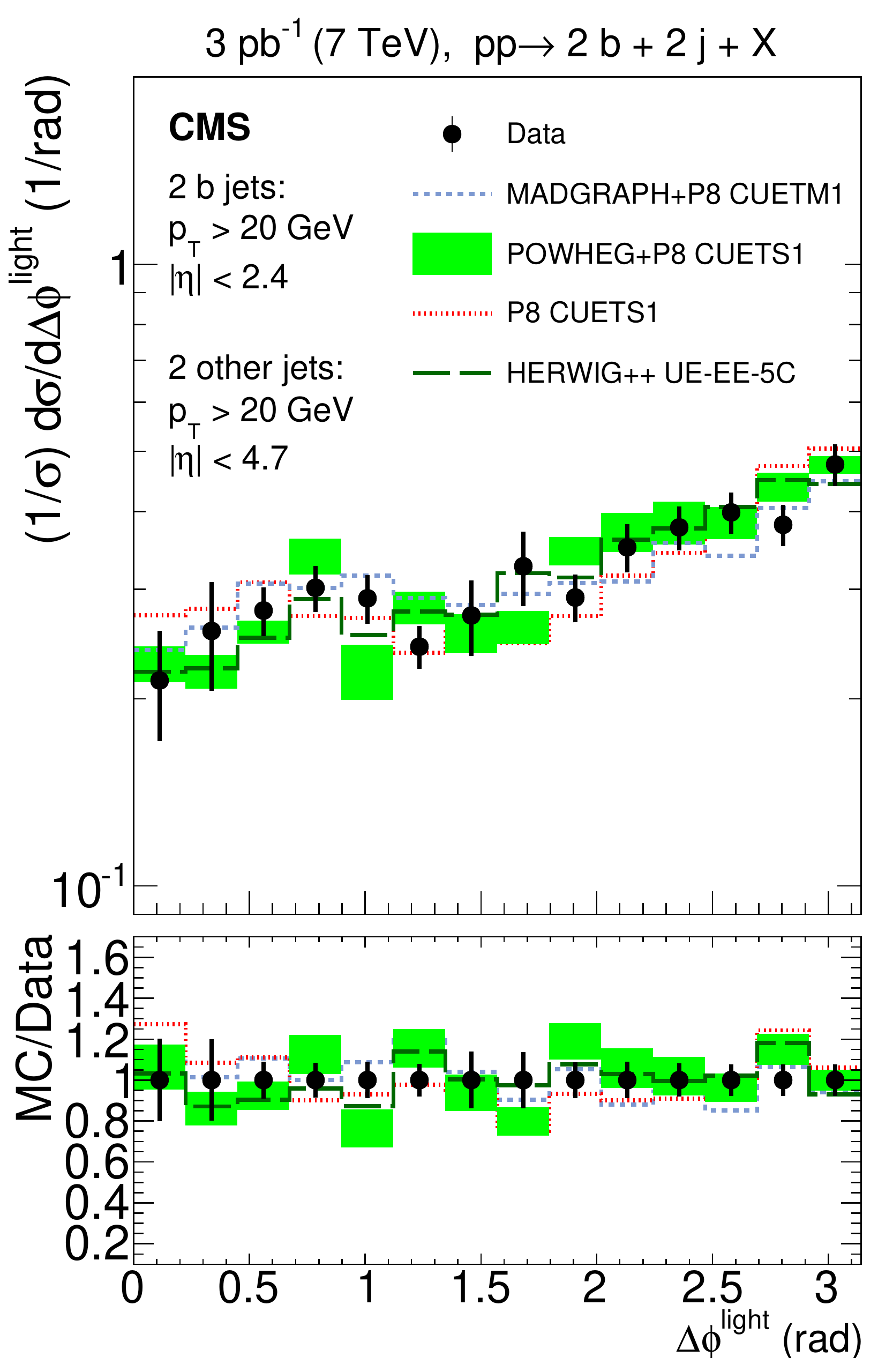}
    \includegraphics[width=\cmsSmallerFigWidth]{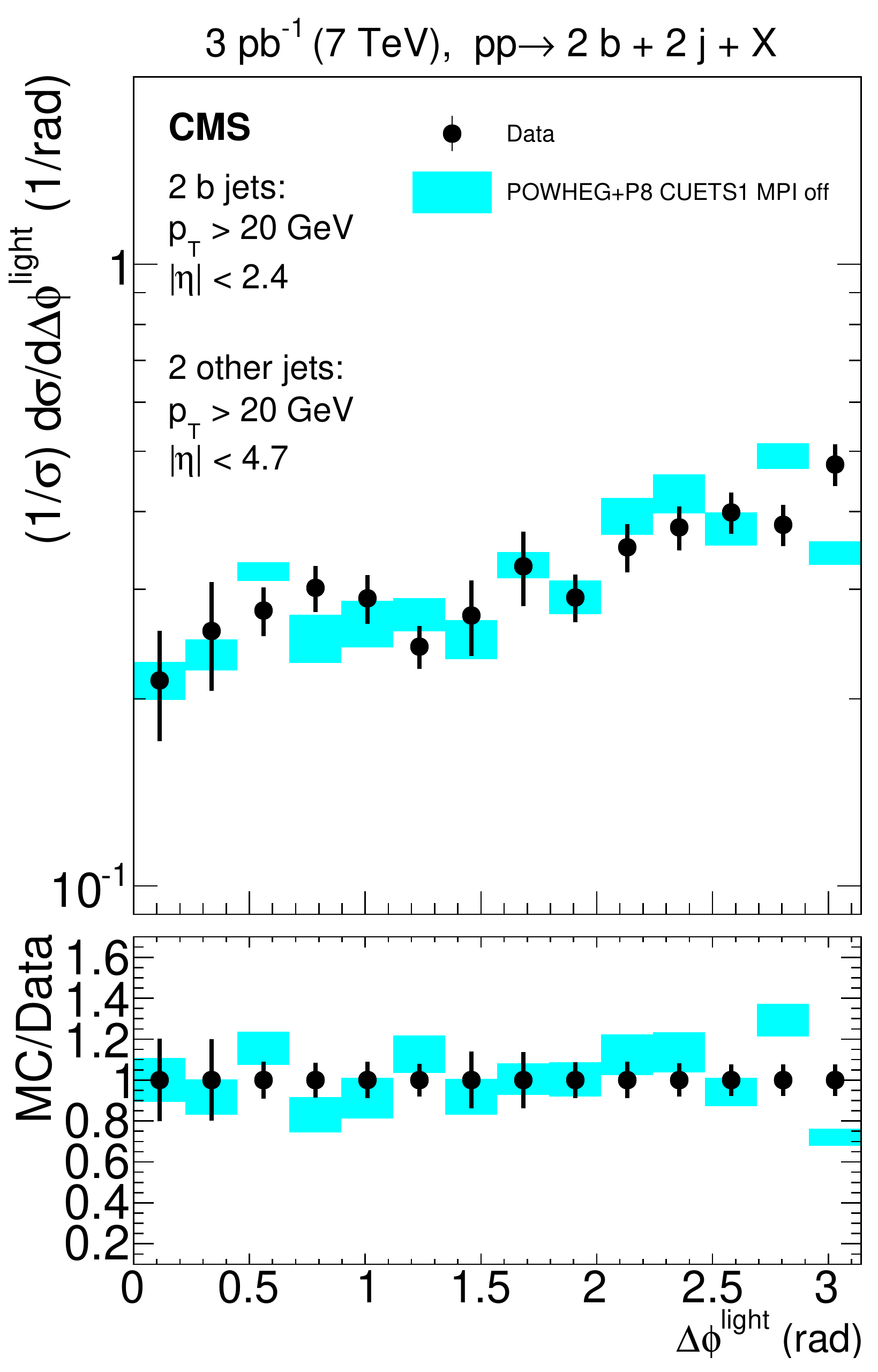}
    \caption{Normalized cross sections unfolded to the particle level as a function of $\Delta\phi^{\text{light}}$, compared to predictions of \POWHEG, \MADGRAPH, \PYTHIA~8 (P8), and \HERWIGpp (\cmsLeft), and of the \POWHEG{}+\PYTHIA~8 tune CUETS1 without MPI (\cmsRight). The lower panels show the ratios of the MC predictions over the data. The error bars on the data represent the total uncertainties, \ie, statistical and systematic added quadratically. Data are shown with markers at unity. The band represents the theoretical uncertainty due to the choice of the scales and PDFs (shown only around the \POWHEG line for clarity, but affecting all predictions in the same way).}
    \label{figShapeCorr1}
\end{figure}

\begin{figure}[htbp]
\centering
    \includegraphics[width=\cmsSmallerFigWidth]{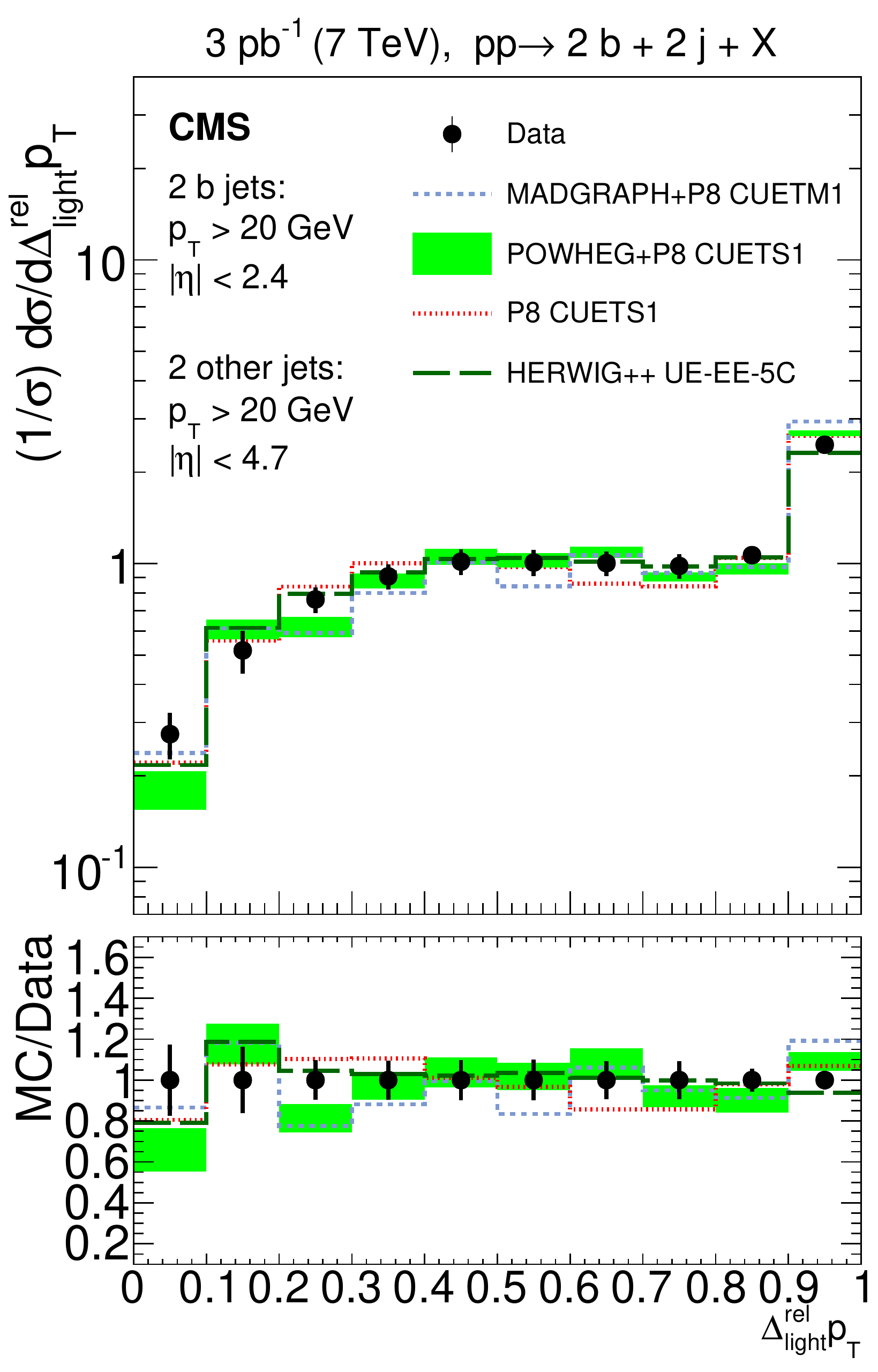}
    \includegraphics[width=\cmsSmallerFigWidth]{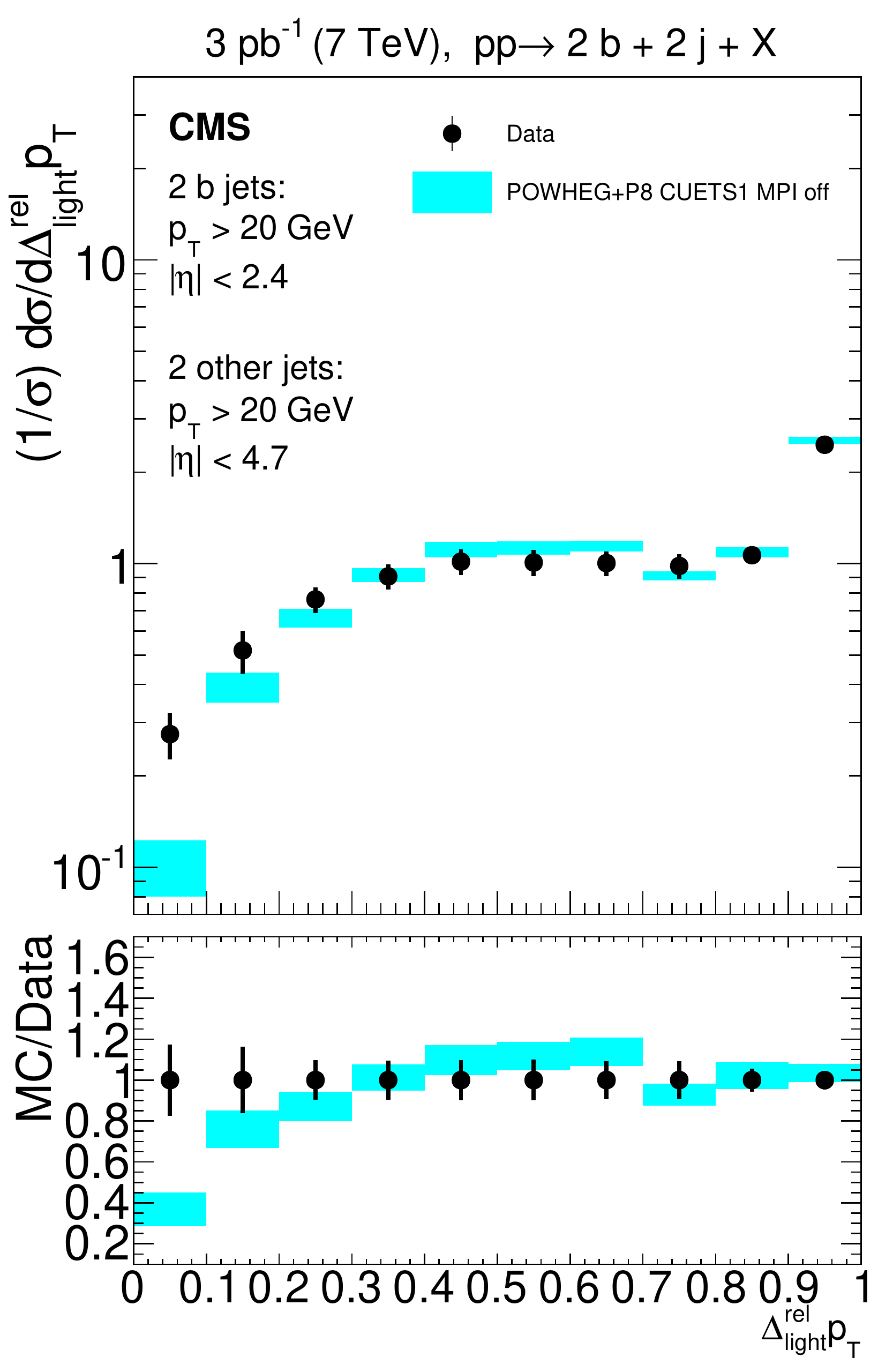}
    \caption{Normalized cross sections unfolded to the particle level as a function of $\Delta^{\text{rel}}_{\text{light}}\pt$, compared to predictions of \POWHEG, \MADGRAPH, \PYTHIA~8 (P8), and \HERWIGpp (\cmsLeft), and of the \POWHEG{}+\PYTHIA~8 tune CUETS1 without MPI (\cmsRight). The lower panels show the ratios of the MC predictions over the data. The error bars on the data represent the total uncertainties, \ie, statistical and systematic added quadratically. Data are shown with markers at unity. The band represents the theoretical uncertainty due to the choice of the scales and PDFs (shown only around the \POWHEG line for clarity, but affecting all predictions in the same way).}
    \label{figShapeCorr2}
\end{figure}

\begin{figure}[htbp]
\centering
    \includegraphics[width=\cmsSmallerFigWidth]{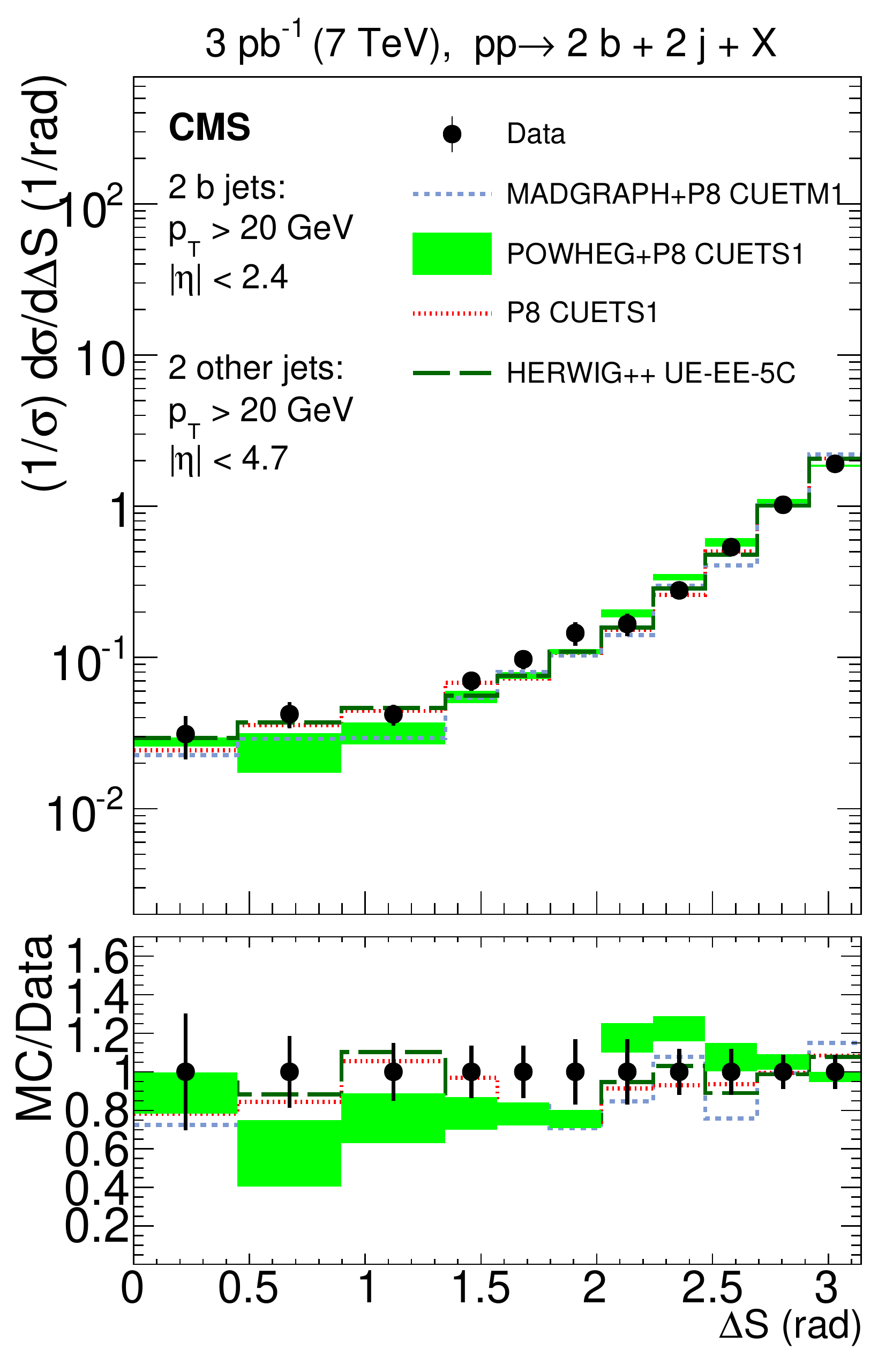}
    \includegraphics[width=\cmsSmallerFigWidth]{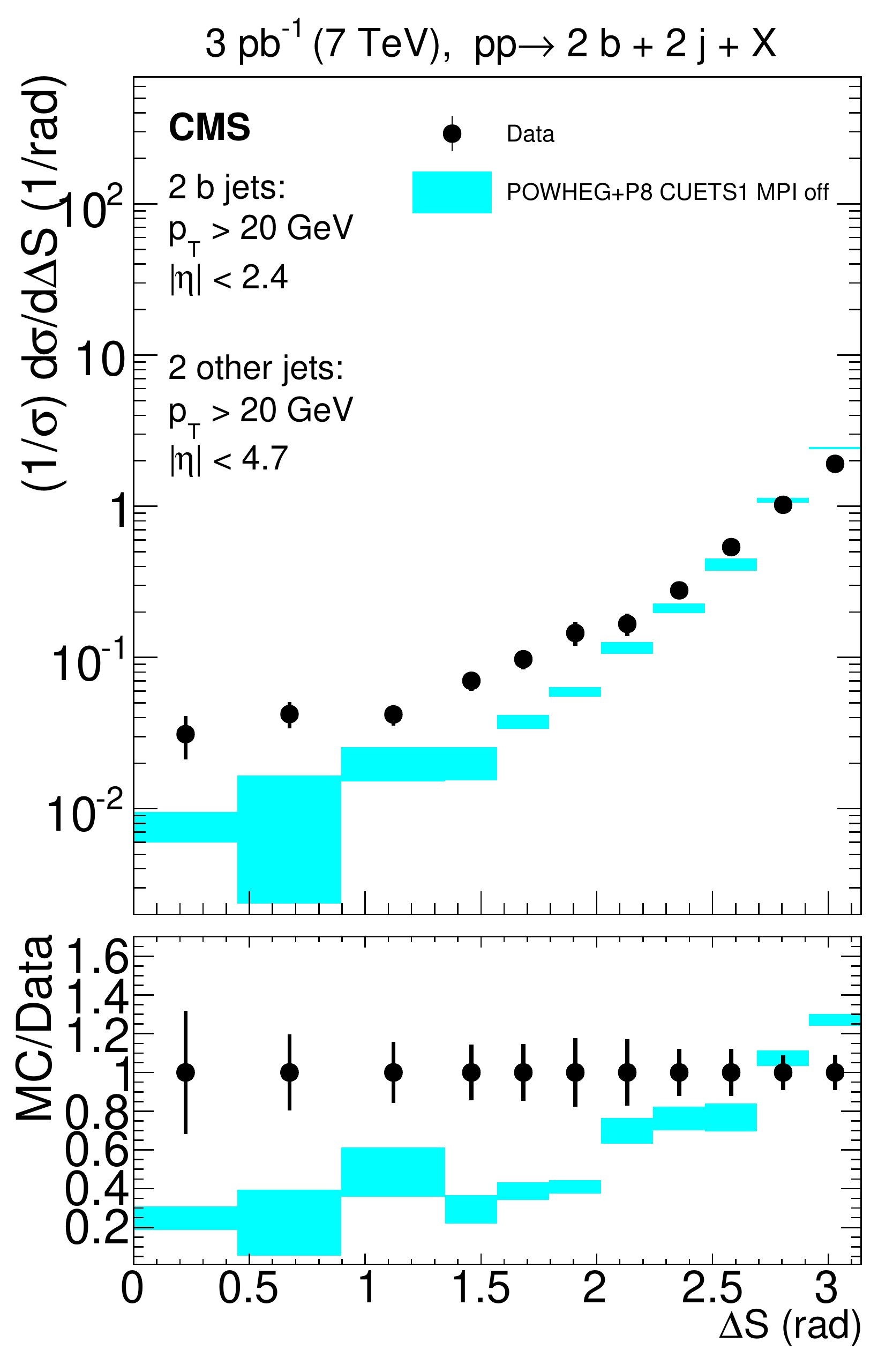}
    \caption{Normalized cross sections unfolded to the particle level as a function of $\Delta$S, compared to predictions of \POWHEG, \MADGRAPH, \PYTHIA~8 (P8), and \HERWIGpp (\cmsLeft), and of the \POWHEG{}+\PYTHIA~8 tune CUETS1 without MPI (\cmsRight). The lower panels show the ratios of the MC predictions over the data. The error bars on the data represent the total uncertainties, \ie, statistical and systematic added quadratically. Data are shown with markers at unity. The band represents the theoretical uncertainty due to the choice of the scales and PDFs (shown only around the \POWHEG line for clarity, but affecting all predictions in the same way).}
    \label{figShapeCorr3}
\end{figure}

Figures \ref{figShapeCorr1}--\ref{figShapeCorr3} show the normalized differential cross sections as a function of the correlation observables, $\Delta\phi^{\text{light}}$, $\Delta^{\text{rel}}_{\text{light}}\pt$, and $\Delta$S. The data are compared to the MC simulations considered previously. In addition, predictions from \POWHEG MPI-off are also shown. All MC simulations that include MPI contributions describe the data well. This is remarkable given that the predictions are based on MPI models tuned to data at softer scales ($\pt \approx3$--5\GeV). Conversely, \POWHEG MPI-off is ruled out by the data, especially at low values of $\Delta^{\text{rel}}_{\text{light}}\pt$ ($<$0.1) and for values of $\Delta$S smaller than 2. This is a clear indication for the need of MPI contributions. The discrepancy between the measurement and the \POWHEG MPI-off predictions goes up to 60\% in the low $\Delta$S region, while for the four-jet events of Ref.~\cite{Chatrchyan:2013qza} the disagreement is of about 40\%. This shows that heavy-flavor multijet production with common jet threshold is more sensitive to a DPS contribution than an untagged four-jet sample with asymmetric \pt thresholds. The fact that the normalized distribution as a function of $\Delta\phi^{\text{light}}$ is also described reasonably well by \POWHEG MPI-off reflects the limited DPS sensitivity of this observable, as already observed for exclusive four-jet final states~\cite{Chatrchyan:2013qza}.

In summary, predictions using LO or NLO dijet matrix elements matched to the simulation of MPI effects reproduce the measured normalized cross sections, whereas those without MPI fail to describe them. This study demonstrates the presence of DPS in the data and confirms the sensitivity to such contributions of the jet correlation variables $\Delta$S and $\Delta^{\text{rel}}_{\text{light}}\pt$.

\section{Summary}
A study of events with at least four jets, at least two of which are \PQb jets, in proton-proton collisions at 7\TeV is presented. The data, corresponding to an integrated luminosity of 3\pbinv, were collected with the CMS experiment in 2010. The two \PQb jets must be within pseudorapidity $\abs{\eta}<2.4$, and the two other jets must be within $\abs{\eta}<4.7$. The transverse momenta of all the jets are required to be greater than 20\GeV. The cross section is measured to be $\sigma(\Pp\Pp \to 2 \PQb + 2 \Pj  + \PX) = 69 \pm 3\stat\pm24\syst\unit{nb}$. The differential cross sections as a function of the \pt and $\eta$ of each of the four jets are presented, along with the cross sections as a function of kinematic jet correlation variables. The results are compared to several theoretical predictions with and without contributions from double parton scattering. The models based on leading order or next-to-leading-order dijet matrix element calculations, matched to parton shower and including multiparton interaction (MPI) contributions, describe well the differential cross sections as a function of \pt and $\eta$ in the whole measured region. The differential cross sections as a function of the jet correlation variables are poorly reproduced by models that do not include contributions from MPI. Specifically, the predictions of \POWHEG interfaced with \PYTHIA~8 without the simulation of multiple parton interactions underestimate the cross sections as a function of $\Delta$S and $\Delta^{\text{rel}}_{\text{light}}\pt$ in the regions of the phase space where a double parton scattering (DPS) signal is expected. These results demonstrate, for the first time, the sensitivity of kinematic jet correlation variables, such as $\Delta$S and $\Delta^{\text{rel}}_{\text{light}}\pt$, to DPS processes in multijet final states with heavy-quarks.

\clearpage
\begin{acknowledgments}
\hyphenation{Rachada-pisek}
We congratulate our colleagues in the CERN accelerator departments for the excellent performance of the LHC and thank the technical and administrative staffs at CERN and at other CMS institutes for their contributions to the success of the CMS effort. In addition, we gratefully acknowledge the computing centers and personnel of the Worldwide LHC Computing Grid for delivering so effectively the computing infrastructure essential to our analyses. Finally, we acknowledge the enduring support for the construction and operation of the LHC and the CMS detector provided by the following funding agencies: BMWFW and FWF (Austria); FNRS and FWO (Belgium); CNPq, CAPES, FAPERJ, and FAPESP (Brazil); MES (Bulgaria); CERN; CAS, MoST, and NSFC (China); COLCIENCIAS (Colombia); MSES and CSF (Croatia); RPF (Cyprus); SENESCYT (Ecuador); MoER, ERC IUT and ERDF (Estonia); Academy of Finland, MEC, and HIP (Finland); CEA and CNRS/IN2P3 (France); BMBF, DFG, and HGF (Germany); GSRT (Greece); OTKA and NIH (Hungary); DAE and DST (India); IPM (Iran); SFI (Ireland); INFN (Italy); MSIP and NRF (Republic of Korea); LAS (Lithuania); MOE and UM (Malaysia); BUAP, CINVESTAV, CONACYT, LNS, SEP, and UASLP-FAI (Mexico); MBIE (New Zealand); PAEC (Pakistan); MSHE and NSC (Poland); FCT (Portugal); JINR (Dubna); MON, RosAtom, RAS and RFBR (Russia); MESTD (Serbia); SEIDI and CPAN (Spain); Swiss Funding Agencies (Switzerland); MST (Taipei); ThEPCenter, IPST, STAR and NSTDA (Thailand); TUBITAK and TAEK (Turkey); NASU and SFFR (Ukraine); STFC (United Kingdom); DOE and NSF (USA).

Individuals have received support from the Marie-Curie program and the European Research Council and EPLANET (European Union); the Leventis Foundation; the A. P. Sloan Foundation; the Alexander von Humboldt Foundation; the Belgian Federal Science Policy Office; the Fonds pour la Formation \`a la Recherche dans l'Industrie et dans l'Agriculture (FRIA-Belgium); the Agentschap voor Innovatie door Wetenschap en Technologie (IWT-Belgium); the Ministry of Education, Youth and Sports (MEYS) of the Czech Republic; the Council of Science and Industrial Research, India; the HOMING PLUS program of the Foundation for Polish Science, cofinanced from European Union, Regional Development Fund, the Mobility Plus program of the Ministry of Science and Higher Education, the National Science Center (Poland), contracts Harmonia 2014/14/M/ST2/00428, Opus 2013/11/B/ST2/04202, 2014/13/B/ST2/02543 and 2014/15/B/ST2/03998, Sonata-bis 2012/07/E/ST2/01406; the Thalis and Aristeia programs cofinanced by EU-ESF and the Greek NSRF; the National Priorities Research Program by Qatar National Research Fund; the Programa Clar\'in-COFUND del Principado de Asturias; the Rachadapisek Sompot Fund for Postdoctoral Fellowship, Chulalongkorn University and the Chulalongkorn Academic into Its 2nd Century Project Advancement Project (Thailand); and the Welch Foundation, contract C-1845. \end{acknowledgments}

\bibliography{auto_generated}

\cleardoublepage \appendix\section{The CMS Collaboration \label{app:collab}}\begin{sloppypar}\hyphenpenalty=5000\widowpenalty=500\clubpenalty=5000\textbf{Yerevan Physics Institute,  Yerevan,  Armenia}\\*[0pt]
V.~Khachatryan, A.M.~Sirunyan, A.~Tumasyan
\vskip\cmsinstskip
\textbf{Institut f\"{u}r Hochenergiephysik der OeAW,  Wien,  Austria}\\*[0pt]
W.~Adam, E.~Asilar, T.~Bergauer, J.~Brandstetter, E.~Brondolin, M.~Dragicevic, J.~Er\"{o}, M.~Flechl, M.~Friedl, R.~Fr\"{u}hwirth\cmsAuthorMark{1}, V.M.~Ghete, C.~Hartl, N.~H\"{o}rmann, J.~Hrubec, M.~Jeitler\cmsAuthorMark{1}, A.~K\"{o}nig, I.~Kr\"{a}tschmer, D.~Liko, T.~Matsushita, I.~Mikulec, D.~Rabady, N.~Rad, B.~Rahbaran, H.~Rohringer, J.~Schieck\cmsAuthorMark{1}, J.~Strauss, W.~Treberer-Treberspurg, W.~Waltenberger, C.-E.~Wulz\cmsAuthorMark{1}
% ...
\vskip\cmsinstskip
% ...
1:~~Also at Vienna University of Technology, Vienna, Austria\\
% ...

\end{sloppypar}
\end{document}